\documentclass[prb,aps,superscriptaddress,showpacs,%
notitlepage,twocolumn,tightenlines,a4paper,%
preprintnumbers]{revtex4}
\usepackage{epsfig}

\newcommand{\bc}{\begin{center}}
\newcommand{\ec}{\end{center}}

\newcommand{\tens}{{\cal T}}

\newcommand{\fr}[2]{{\frac{#1}{#2}}}

\newcommand{\be}{\begin{equation}}
\newcommand{\ee}{\end{equation}}
\newcommand{\ba}{\begin{eqnarray}}
\newcommand{\ea}{\end{eqnarray}}

\newcommand{\bi}{\begin{itemize}}
\newcommand{\ei}{\end{itemize}}

\newcommand{\nr}[1]{(\ref{#1})}

\renewcommand{\vec}[1]{{\bf #1}}
\newcommand{\twovec}[1]{\underline{\bf #1}}

\newcommand{\eq}[1]{Eq.\ (\ref{#1})}
\newcommand{\peq}[1]{Eq.\ (\ref{#1})}

\newcommand{\plaq}{P}
\newcommand{\fzs}{{\rm FZS}}
\newcommand{\xy}{{\rm XY}}
\newcommand{\chimexp}{\upsilon}
\newcommand{\tensexp}{\nu_\tens}

\def\lsi{\raise0.3ex\hbox{$<$\kern-0.75em\raise-1.1ex\hbox{$\sim$}}}
\def\gsi{\raise0.3ex\hbox{$>$\kern-0.75em\raise-1.1ex\hbox{$\sim$}}}
\newcommand{\lsim}{\mathop{\lsi}}

\begin{document}

\title{Numerical study of duality and
 universality in a frozen superconductor}
\author{T. Neuhaus}
\email{neuhaus@pcu.helsinki.fi}
\affiliation{Finkenweg 15, D-33824 Werther, Germany}

\author{A. Rajantie}
\email{a.k.rajantie@damtp.cam.ac.uk}
\affiliation{DAMTP, CMS, University of Cambridge, 
Cambridge CB3 0WA, United Kingdom}
\affiliation{Institute for Theoretical Physics,
University of California, Santa Barbara, CA 93106, USA
}

\author{K. Rummukainen}
\email{kari@nordita.dk}
\affiliation{NORDITA, Blegdamsvej 17, 
DK-2100 Copenhagen \O, Denmark}
\affiliation{Department of Physics, P.O.Box 64, FIN-00014 
University of Helsinki, Finland}

\date{14 January, 2003}

\begin{abstract}
The three-dimensional
integer-valued lattice gauge theory, which is also known as a ``frozen
superconductor,'' can be obtained as a certain limit of the
Ginzburg-Landau theory of superconductivity, and is believed to be in
the same universality class. It is also exactly dual
to the three-dimensional XY model. We use this duality to 
demonstrate the practicality of recently developed methods for
studying topological defects, and investigate the critical behaviour
of the phase transition using numerical Monte Carlo
simulations of both theories. On the gauge theory side, we concentrate
on the vortex tension and the penetration 
depth,
which map onto the
correlation lengths of the order parameter and the Noether current in
the XY model, respectively. 
We show how these quantities behave near the critical
point, and that the penetration 
depth
 exhibits critical 
scaling only very close to the transition point. This may explain the
failure of superconductor experiments to see the inverted XY model scaling.
\end{abstract}

\preprint{NORDITA-2002/31 HE}
\preprint{DAMTP-2002-55}
\preprint{NSF-ITP-02-42}


\maketitle

\section{Introduction}
Duality arguments 
suggest that the 
superconductor-insulator 
phase transition should be in the same universality class as the
three-dimensional XY model,
but with an inverted temperature
axis.\cite{Peskin:1978kp,dasgupta1981,%
Kovner:nk,kiometzis1994,kiometzis1995} 
This means that the superconducting phase maps onto the
symmetric phase, and the normal phase onto the broken
phase. Furthermore, the dual counterpart of the order parameter of the
XY model is a non-perturbative field that creates the
Abrikosov-Nielsen-Olesen vortices in the superconductor.

The most direct prediction of the duality is that the critical
magnetic field $H_{\rm c1}$, or equivalently the vortex tension $\tens$,
should scale with the XY model critical exponent $\nu_{\rm XY}\approx
0.6723$.\cite{Hasenbusch:1999cc}
In practice, it is easier to measure the scaling exponent $\nu'$ 
of the penetration 
depth
 $\lambda$,
but for that quantity, the theoretical picture is less clear.
Both $\nu'\approx 0.33$~\cite{fisher1991} and 
$\nu'\approx 0.5$~\cite{kiometzis1994,kiometzis1995} were suggested before 
the prediction eventually converged to
$\nu'=\nu_{\rm XY}$.\cite{herbut1996}

Ironically, two different experiments with
YBa$_2$Cu$_3$O$_{7-\delta}$ high-temperature superconductor have
produced results that are each compatible with one of the earlier
suggestions, namely
$\nu'=0.34(1)$,\cite{kamal1998} and
$\nu'=0.45\ldots 0.5$.\cite{paget1999}

Similarly, it has turned out to be difficult to confirm the duality
in Monte Carlo simulations of the Ginzburg-Landau theory away from the
London limit,\cite{Kajantie:2001ts} as they also seem to favour
$\nu'\approx 0.3$. In the London limit, where the duality is on a
firmer footing,\cite{Thomas:1978aa,Kleinert:dz}
simulations~\cite{olsson1998,hove2000} give
$\nu'\approx 0.67$ through indirect measurements, though. 

There are two principal reasons for all this confusion:

\begin{enumerate}
\item The duality relation is expected to apply
quantitatively only in a narrow temperature interval near the
critical point, where the vortex tension/$k_BT$ is well below the
inverse correlation lengths of the scalar and photon fields.  
\item
The duality relates the fundamental fields and thermodynamical
densities of one theory to non-perturbative and non-local objects in
the other.   These are difficult to measure both in experiments
and in numerical simulations.
\end{enumerate}

In the London limit, the difficulty (1) is alleviated, as the scalar
correlation length is very short, but in this paper we go even further
and take another limit to obtain the Abelian integer-valued lattice
gauge theory, ``frozen superconductor'' (\fzs).\cite{Peskin:1978kp} 
For this theory, the duality transformation can
be carried out exactly, and yields precisely the three-dimensional XY
model with the Villain
action.\cite{Banks:1977cc,Savit:1977ag,Thomas:1978aa,Peskin:1978kp}
The duality is therefore exactly valid at all temperatures.

The \fzs\ does not have a scalar (Higgs) field to drive the
transition, but it still has a transition between a low-temperature
superconducting phase with massive gauge field excitations, and a
high-temperature massless phase. 
(Microscopically, this
transition is due to the ``freezing'' of the discrete gauge variables
below the critical temperature; hence the name frozen superconductor.)
This is mapped to the symmetry
breaking transition of the XY-Villain model, with inverted temperature
($T\leftrightarrow 1/T$).  Since the duality is valid at all
temperatures, the problem (1) above is completely avoided.

In this paper we study the duality relation in detail with lattice
Monte Carlo simulations.  Our aim is {\em not} to numerically verify
the duality --- it is, after all, a mathematical identity.  Rather,
the purpose is to identify ``good'' observables and possible pitfalls
in Monte Carlo studies of the duality relations of this
type,  especially bearing in mind the problems encountered in the GL theory
simulations. 
By this we mean, first of all, that we want to construct pairs of 
observables in the frozen superconductor and in the XY model
which are exactly dual to each other and which are sensitive
to the critical behaviour.  The observable in the 
frozen superconductor must be defined also in the full GL model.
It is, however, equally important that the observables
can be measured to a high accuracy in both theories.
The exact duality of the models we consider allows us 
to check this and
identify the errors caused by finite-size effects
and possible inefficiencies caused by the numerical methods used.
In this way, we hope to find out how to avoid the
difficulty (2) in more realistic cases.

The main body of the paper is concerned with the
calculation of gauge invariant (non-local) order parameters of
the \fzs; the vortex tension $\tens$, the photon
mass $m_\gamma$ and the magnetic permeability
$\chi_m$.  We relate these observables to their duals
in the XY model, and study the critical behaviour.

This paper is organized 
as follows. Section~2 contains the definition of the two models 
and discusses the duality between them. Section~3 introduces the observables 
and Section 4~discusses their qualitative behaviour in the two
phases. In Section~5, the location of
critical point is determined using the spin correlator in the XY
model, and
Section~6 deals with the definition and the measurement of its dual
counterpart, the vortex tension.
In Section~8 we study the susceptibilities, i.e., the photon 
correlator in the momentum space, and discuss the mass determination
from them, and in
Section~7, we extract the photon mass directly from the exponential
decay of the 
correlation function in the coordinate space.
Finally, Section~9 lists the main findings and 
concludes the paper.

\section{Duality}
\label{sect:dual}

We start by formulating the three-dimensional Ginzburg-Landau theory
in the London limit 
on a three-dimensional cubic lattice.
The theory consists of a real valued gauge
field $A_{\vec{x},i}$ defined on links and spin angles
$\theta_\vec{x}$ defined on lattice sites. The partition function is
\begin{equation}
Z_{\rm GL}=\int \left(\prod_\vec{x} d\theta_\vec{x} \prod_i
dA_{\vec{x},i}
\right)\exp\left(-\sum_\vec{x}{\cal L}_{\rm GL,\vec{x}}\right),
\label{equ:glpart}
\end{equation}
where
\begin{equation}
{\cal L}_{\rm GL,\vec{x}}=
-\frac{1}{2}\sum_{i<j}F_{\vec{x},ij}^2
+\kappa\sum_i s\left(\theta_{\vec{x}+i}-\theta_\vec{x}-qA_{\vec{x},i}
\right),
\label{equ:gllagr}
\end{equation}
and
\begin{equation}
F_{\vec{x},ij}= A_{\vec{x},i}+A_{\vec{x}+i,j}-A_{\vec{x}+j,i}-
A_{\vec{x},j},
\end{equation}
and $s(x)$ is a periodic function with period $2\pi$ and minimum at $x=0$.
The standard choice for $s(x)$ is $s(x)=-\cos(x)$, but here we shall use
the Villain form\cite{Villain:1975ir}
\begin{equation}
s(x) = -{\rm ln} \sum_{k=-\infty}^{\infty} 
\exp\left(- \frac{1}{2} (x-2\pi k)^2\right).
\label{equ:Villain}
\end{equation}
In this paper we shall consider two limits of this theory. The frozen
superconductor (\fzs),\cite{Thomas:1978aa,Peskin:1978kp} which
we also refer to as simply ``the gauge theory,'' 
is obtained by taking $\kappa\rightarrow\infty$ and
defining $\beta=4\pi^2/q^2$. This leads to the partition function
\begin{equation}
Z_{\fzs}(\beta)=\sum_{\{I_{\vec{x},i}\}} 
\exp\left(- \frac{\beta}{2} \sum_{\vec{x},i>j} \plaq_{\vec{x},ij}^2\right)\,,
\label{iz-partition-function}
\end{equation}
where
\begin{equation}
\label{equ:plaqint}
\plaq_{\vec{x},ij}= I_{\vec{x},i}+I_{\vec{x}+i,j}-I_{\vec{x}+j,i}-
I_{\vec{x},j},
\end{equation}
and the link variables $I_{\vec{x},i}=qA_{\vec{x},i}/2\pi$ take
integer values. This model has two phases: the Coulomb phase
($\beta<\beta_c$) and the superconducting phase ($\beta>\beta_c$).

We also consider the limit $q\rightarrow 0$, in which we recover the 
three-dimensional XY model. The gauge field
$A_{\vec{x},i}$ decouples, and the non-trivial spin part of the partition
function becomes
\begin{equation}
Z_{\xy}(\kappa)=\int {D\theta} \exp\left(-\kappa \sum_{\vec{x},i} 
s(\theta_{\vec{x}+i}-\theta_\vec{x})\right).
\label{xy-partition-function}
\end{equation}
This model has a broken phase at $\kappa>\kappa_c$ and a symmetric
phase at $\kappa<\kappa_c$.

\begin{widetext}
We shall now show that the partition functions
(\ref{iz-partition-function}) and (\ref{xy-partition-function}) are
dual to each other, i.e.,
proportional to each other if $\beta=1/\kappa$.
Introducing a real vector field $h_{\vec{x},i}$,\cite{Jose:1977gm}
we can write
\eq{xy-partition-function} as
\begin{equation}
Z_{\xy}(\kappa)\propto\int {D\theta}Dh_i\sum_{k_i}
\exp\left[
-\sum_{\vec{x},i}\left(
\frac{1}{2\kappa}h_{\vec{x},i}^2-ih_{\vec{x},i}\Delta^{\xy}_{\vec{x},i}
\right)
\right],
\label{equ:Zxy}
\end{equation}
where we have introduced the Noether current of the XY model
\begin{equation}
\Delta^{\xy}_{\vec{x},i}=\theta_{\vec{x}+i}-\theta_{\vec{x}}-2\pi
k_{\vec{x},i}.
\label{equ:deltadef}
\end{equation}
The integration over $\theta$ yields a delta function
$\delta(\vec{\nabla}\cdot\vec{h})$, 
where we have defined the lattice divergence
\begin{equation}
\vec{\nabla}\cdot\vec{h}_\vec{x}=\sum_i\left(h_{\vec{x},i}-h_{\vec{x}-i,i}
\right).
\end{equation}
The summation over
$k_{\vec{x},i}$ restricts $h_{\vec{x},i}$ to integer values, and we
obtain\cite{Banks:1977cc}
\begin{equation}
Z_{\xy}(\kappa)\propto\sum_{\{h\}}\delta_{\vec{\nabla}\cdot\vec{h},0}
\exp\left(-\frac{1}{2\kappa}\sum_{\vec{x},i}h_{\vec{x},i}^2\right),
\label{equ:sumh}
\end{equation}
which is the partition function of an integer-valued and sourceless
vector field.  In an infinite volume, 
we can interpret the vector field
$h_{\vec{x},i}$ as the integer valued flux through the dual lattice
plaquette pierced by link $(\vec{x},i)$, and write
\begin{equation}
h_{\vec{x},i}=\frac{1}{2}\epsilon_{ijk}\plaq_{\vec{x},jk},
\label{equ:hplaq}
\end{equation}
with $\plaq_{\vec{x},jk}$ as in Eq.~(\ref{equ:plaqint}).

Thus, 
identifying $\beta=1/\kappa$, we recover the partition function in
\eq{iz-partition-function}.\cite{Savit:1977ag,Thomas:1978aa,Peskin:1978kp}
This shows that the two limits of 
Eq.~(\ref{equ:glpart}), $\kappa\rightarrow\infty$ and
$q\rightarrow 0$, are dual to each other.
In particular, the duality relation implies that 
the gauge theory has a phase transition
of the XY model universality class at $\beta_c =
1/\kappa_c$, and the Coulomb and superconducting phases are mapped to
the broken and symmetric phases, respectively. 
Note, however, that duality only maps the two limits onto each other; 
the Ginzburg-Landau theory is not self-dual for
finite parameter values.

The nature of the duality transformation becomes more transparent when
we introduce an external, real-valued vector
field $\alpha_{\vec{x},i}$ in the XY model
\begin{equation}
Z_{\xy}\left(\kappa;\{\alpha_{\vec{x},i}\}\right)
=\int {D\theta}\sum_{\{k_{\vec{x},i}\}} 
\exp\left[- \sum_{\vec{x},i} \left(
\frac{\kappa}{2}\left(
\Delta^{\xy}_{\vec{x},i}\right)^2
+i\alpha_{\vec{x},i}\Delta^{\xy}_{\vec{x},i}
\right)\right].
\label{equ:Zalpha}
\end{equation}
Introducing $h_{\vec{x},i}$ as in Eq.~(\ref{equ:Zxy}), but defining
$\tilde{h}_{\vec{x},i}=h_{\vec{x},i}-\alpha_{\vec{x},i}$, we obtain
\begin{equation}
Z_{\xy}(\kappa;\{\alpha_{\vec{x},i}\})\propto\int {D\theta}D\tilde{h}_i\sum_{k_i}
\exp\left[
-\sum_{\vec{x},i}\left(
\frac{1}{2\kappa}\left(\tilde{h}_{\vec{x},i}+\alpha_{\vec{x},i}\right)^2-i
\tilde{h}_{\vec{x},i}
\Delta^{\xy}_{\vec{x},i}
\right)
\right].
\label{equ:Zxyalpha}
\end{equation}
Now, as before, we integrate over $\theta_\vec{x}$, 
which constrains $\vec{\nabla}\cdot\tilde{\vec{h}}_\vec{x}=0$,
sum over $k_{\vec{x},i}$, which makes $\tilde{h}_{\vec{x},i}$ an
integer, and express
$\tilde{h}_{\vec{x},i}$ in terms of $\plaq_{\vec{x},ij}$ as in
Eq.~(\ref{equ:hplaq}). This gives us
\begin{equation}
Z_{\xy}\left(\kappa;\{\alpha_{\vec{x},i}\}\right)=
Z_{\fzs}\left(1/\kappa;\{\alpha_{\vec{x},i}\}\right),
\label{equ:duality}
\end{equation}
 where 
\begin{equation}
Z_{\fzs}\left(\beta;\{\alpha_{\vec{x},i}\}\right)
=\sum_{\{I_{\vec{x},i}\}} 
\exp\left[- \frac{\beta}{2} \sum_{\vec{x},i} 
\left( \fr12 \epsilon_{ijk}  \plaq_{\vec{x},jk}+\alpha_{\vec{x},i}\right)^2\right],
\label{iz-with-alpha}
\end{equation}
Note that in this expression the vector field $\alpha$ is defined
on the dual lattice, i.e. $\alpha_{\vec{x},i}$ lives on the
link that pierces the plaquette $\plaq_{\vec{x},jk}$.
Eq.~(\ref{equ:duality}) is the basic duality equation, which can be
used to relate the observables of the two models to each other.
\end{widetext}

\section{Observables}
\label{equ:observables}

\subsection{Spin-spin correlator}
The basic observable in the XY model is the spin-spin correlation
function,
\begin{equation}
G_{\vec{x}_1,\vec{x}_2}\equiv\left\langle\exp[ i(\theta_{\vec{x}_1}\!-\!
\theta_{\vec{x}_2})]\right\rangle_{\xy}.
\label{equ:xycorr}
\end{equation}
Using Eq.~(\ref{equ:Zalpha}), we can write this as
\begin{equation}
G_{\vec{x}_1,\vec{x}_2}=
\frac{Z_{\xy}\left(\kappa;\{\alpha_{\vec{x},i}\}\right)}
{Z_{\xy}(\kappa)},
\label{equ:introalpha}
\end{equation}
where
$\alpha_{\vec{x},i}$ is an otherwise arbitrary fixed 
integer-valued vector field, 
but it satisfies the condition
\begin{equation}
\vec{\nabla}\cdot\vec{\alpha} =
\delta_{\vec{x},\vec{x}_1}-
\delta_{\vec{x},\vec{x}_2},
\label{equ:sourcesink}
\end{equation}
i.e. it has a source and a ``sink'' at points $\vec{x}_1$ and $\vec{x}_2$,
respectively.  

The duality relation in
Eq.~(\ref{equ:duality}) implies that Eq.~(\ref{equ:introalpha}) must be
equal to
\begin{equation}
G_{\vec{x}_1,\vec{x}_2}
=\left\langle \exp\left(- \frac{\beta}{2}  \sum_{\vec{x},i} 
\left(
\epsilon_{ijk} \alpha_{\vec{x},i}\plaq_{\vec{x},jk}+\alpha_{\vec{x},i}^2
\right)\right)
\right\rangle_{\fzs}.
\label{equ:dualcorr}
\end{equation}
The simplest choice for $\alpha_{\vec{x},i}$ 
vanishes everywhere except on the shortest path of
links that leads from $\vec{x}_1$ to $\vec{x}_2$, on which it has the
value of unity. As we shall see later in Sec.~\ref{sec:tension}, 
\eq{equ:dualcorr} then has 
a natural interpretation as a vortex correlation function 
in the gauge theory.\cite{Banks:1977cc,Peskin:1978kp,Kajantie:1999zn}
In Table~\ref{tab:observables} we summarize the duality between basic 
observables in the XY model and in the frozen superconductor.

In the symmetric phase of the XY model, i.e., when $\kappa<\kappa_c$, 
the spin-spin correlator decays exponentially as
\begin{equation}
G_{\vec{x}_1,\vec{x}_2} \propto e^{-m \mid \vec{x}_1-\vec{x}_2 \mid },
\end{equation}
where we call the decay rate $m$ the scalar mass, in accordance with a
field theory picture.
By definition, the correlation length is given by its inverse
$\xi=1/m$.
Eq.~(\ref{equ:dualcorr}) implies that under the duality
transformation, the scalar mass becomes the
vortex tension 
defined in Sec.~\ref{sec:tension},
$\tens=m$.

In the broken phase of the XY model, where $\kappa>\kappa_c$, 
the spin-spin correlator~(\ref{equ:xycorr}) approaches a
constant
\begin{equation}
\lim_{|\vec{x}_1-\vec{x}_2|\rightarrow\infty}
G_{\vec{x}_1,\vec{x}_2}
 = M^2, 
\end{equation}
where $M$ is the magnetization.

More generally one can also consider higher $n$-point functions in the
XY model, which correspond to external fields $\alpha_{\vec{x},i}$
with more sources and sinks.

We would like to emphasize the difference between our approach and
the
attempts to describe the phase transition of the XY model as vortex
percolation.\cite{pochinsky1991,Hulsebos:1994mn,nguyen1999}
The ``line tension'' discussed in that context 
is defined using the length distribution of vortices in
the XY model and suffers from certain ambiguities,
and there is numerical evidence that the percolation point at which 
the line tension vanishes does not even coincide with the
thermodynamic critical point $\kappa_c$.\cite{Kajantie:2000cw}
In contrast,
our $\tens$ is the vortex tension in the gauge
theory, is a well-defined, unambiguous observable and, 
due to the exact nature of the duality, reflects the true
thermodynamic properties of the system.

\subsection{Helicity modulus}

Another important quantity in the XY model is the helicity 
modulus,\cite{fisher1973}
which characterizes the response of the free energy to a twist of the
spins by an amount $\delta\theta$ along, say, the $z$ direction.
Instead of period boundary conditions for $\theta$, we would instead
have $\theta(x,y,z+N)=\theta(x,y,z)+\delta\theta$. It is convenient
to define a periodic variable
\begin{equation}
\tilde{\theta}(x,y,z)=\theta(x,y,z)-\frac{z}{N_3}\delta\theta\equiv
\theta(x,y,z)-zj_3,
\end{equation}
where $j_3=\delta\theta/N$ is the average Noether current created by
the twist. We can then write the twisted partition function as
\begin{equation}
Z_{\xy}\left(\kappa,j_i\right)
=\int {D\theta}\sum_{\{k_{\vec{x},i}\}} 
\exp\left[- \sum_{\vec{x},i} \left(
\frac{\kappa}{2}\left(
\Delta^{\xy}_{\vec{x},i}+j_i\right)^2
\right)\right],
\label{equ:Zj0}
\end{equation}
where $\vec{j}=(0,0,j_3)$.

The helicity modulus is defined as
\begin{equation}
\Upsilon=\frac{N_3}{N_1N_2}
\left( \frac{\partial^2 F}{\partial \delta\theta^2} \right)_{ \delta\theta=0}
=\frac{1}{V} 
\left( \frac {\partial^2 F} {\partial j_3^2} \right)_{j_3=0},
\end{equation}
where $F=-\ln Z_{\xy}$ is the free energy of the system. 
It is an order parameter, 
with finite, non-zero 
values $\Upsilon>0$ in the broken phase ($\kappa>\kappa_c$)
and a vanishing value $\Upsilon=0$ in the symmetric phase
($\kappa<\kappa_c$).  
If the XY model is viewed as a model for a superfluid, $\Upsilon$ is 
proportional to the superfluid density, $\Upsilon\propto\rho_s$.
We can also write $\Upsilon$ as
\begin{eqnarray}
\Upsilon&\!=\!&-\frac{1}{V}
\frac{1}{Z_{\xy}}\frac{\partial^2 Z_{\xy}}{\partial j_3^2}
=-\frac{\kappa}{V}\left[
\kappa\left\langle\left(\sum_\vec{x}\Delta^{\xy}_{\vec{x},3}
\right)^2\right\rangle-V\right]
\nonumber\\
&\!=\!&
\kappa\left(1-\kappa\chi\right),
\label{equ:upsilon-chi}
\end{eqnarray}
where $\chi=\sum_\vec{x}\langle \Delta^{\xy}_{\vec{0},3}\Delta^{\xy}_{\vec{x},3}
\rangle$ is a
susceptibility related to the U(1) current density of the XY model.

It is interesting to ask what the helicity modulus 
$\Upsilon$ corresponds to in the gauge
theory.
Formally, it is related to
Eq.~(\ref{equ:Zalpha}) with a constant imaginary field
$\alpha_i=-i\kappa j_i$,
\begin{equation}
Z_{\xy}\left(\kappa,j_i\right)=
Z_{\xy}\left(\kappa;\{\alpha_i\}\right)\exp\left(
-\frac{\kappa}{2}Vj_i^2
\right),
\end{equation}
and therefore 
\begin{equation}
\Upsilon = \frac{\kappa^2}{V}\frac{1}{Z_{\xy}}
\frac{\partial^2 Z_{\xy}}{\partial \alpha_3^2}+\kappa
=
\frac{1}{\beta^2V}\frac{1}{Z_{\fzs}}
\frac{\partial^2 Z_{\fzs}}{\partial \alpha_3^2}+\frac{1}{\beta}.
\label{equ:upsilon-in-iz}
\end{equation}
Using Eq.~(\ref{iz-partition-function}), this is nothing but
\begin{equation}
\Upsilon = \frac{1}{V}\left\langle\left(\sum_\vec{x}\plaq_{\vec{x},12}
\right)^2\right\rangle\equiv \chi_m,
\label{equ:ups-equals-chim}
\end{equation}
where we have defined the magnetic permeability $\chi_m$
as the susceptibility associated with the magnetic
flux in the \fzs.

\begin{table*}
\begin{tabular}{ll|ll}
&XY model & \fzs &\\
\hline
spin-spin correlator & $G_{\vec{x}_1,\vec{x}_2}$ & 
Eq.~(\ref{equ:dualcorr}) & vortex correlator\\
scalar mass & $m=1/\xi$ & $\tens$ & vortex tension\\
helicity modulus & $\Upsilon$ & $\chi_m$ & magnetic permeability\\
current-current correlator & $-\beta^{-2}\langle
\Delta^{\xy}_{\vec{x}_1,i}\Delta^{\xy}_{\vec{x}_2,j}
\rangle$ & $\Gamma_{(\vec{x}_1,i)(\vec{x}_2,j)}$ & photon correlator \\
\hline
\end{tabular}

\caption{
Comparison of the basic observables in the XY model and in the frozen
superconductor (FZS).
\label{tab:observables}}
\end{table*}

\subsection{Photon mass}

{}From the \fzs\ point of view, the most natural observable
is the photon correlator, defined as the correlation function of
plaquettes,
\begin{equation}
\Gamma_{(\vec{x}_1,i)(\vec{x}_2,j)}=\langle \plaq_{\vec{x}_1,kl} 
\plaq_{\vec{x}_2,mn}\rangle_\fzs .\quad
(ikl~\mbox{and}~jmn~\mbox{cyclic})
\label{equ:photoncorr}
\end{equation}
In the superconducting phase ($\beta>\beta_c$), this correlator decays
exponentially, and we call the decay rate the photon mass $m_\gamma$.
The penetration
depth
 $\lambda$ 
is defined as the inverse of the photon mass,
$\lambda=1/m_\gamma$.
In the non-superconducting Coulomb phase ($\beta < \beta_c$), $\Gamma_{(\vec{x}_1,i)(\vec{x}_2,j)}$
has a power-law decay, which corresponds to vanishing
photon mass $m_\gamma=0$, or infinite penetration
depth
 $\lambda=\infty$.

We write the photon correlation function as
\begin{equation}
\Gamma_{(\vec{x}_1,i)(\vec{x}_2,j)}=\left.
\frac{1}{\beta^2 Z_{\fzs}(\beta)}
\frac{\partial^2 Z_{\fzs}\left(\beta;\{\alpha_{\vec{x},i}\}\right)
}{\partial \alpha_{\vec{x}_1,i}\partial \alpha_{\vec{x}_2,j}}
\right|_{{\alpha=0}}
\end{equation}
which is simply a generalization of Eq.~(\ref{equ:upsilon-in-iz}) to a
non-constant $\alpha_{\vec{x},i}$. Using the duality
(\ref{equ:duality}), we find
\begin{equation}
\Gamma_{(\vec{x},i)(\vec{y},j)}=
\frac{1}{\beta}\delta_{\vec{x},\vec{y}}\delta_{ij}
-\frac{1}{\beta^2}\left\langle
\Delta^{\xy}_{\vec{x},i}\Delta^{\xy}_{\vec{y},j}
\right\rangle_\xy.
\label{eq:mapped_photon}
\end{equation}

In practice, it is more convenient to study correlations between
planes rather than individual plaquettes. To this end, we interpret the $z$ direction as
``time'' and label it with $\tau$. We consider the photon correlation function
with a non-zero ``spatial'' momentum $\twovec{p}$,
where we use the underline to indicate a two-vector in the
$(1,2)$ 
plane.  (The zero-momentum correlation function vanishes identically.)
The correlation function is defined as
\begin{equation}
\Gamma(\tau,\twovec{p})=
\frac{1}{N_3}{\rm Re} \left\lbrace \sum_{\tau_0}\sum_{\twovec{x},\twovec{y}}
\langle \plaq_{\vec{x},12}\plaq_{\vec{y},12}\rangle
e^{-i \twovec{p}\cdot (\twovec{x}-\twovec{y})} \right\rbrace,
\label{photoncorrelation}
\end{equation}
where $p_i=2\pi k_i/N_i$, $(i=1,2)$ and $k_i$ are integers,
$\vec{x}=(\tau_0,\twovec{x})$ and $\vec{y}=(\tau_0+\tau,\twovec{y})$.

It is also useful to consider the three-dimensional 
Fourier transform of the photon
correlator
\begin{equation}
\Gamma_{ij}(\vec{p})=
\sum_\vec{x} e^{-i\vec{p}\cdot\vec{x}}
\Gamma_{(\vec{0},i)(\vec{x},j)}.
\label{equ:photoncorr-fourier}
\end{equation}
Eq.~(\ref{eq:mapped_photon}) then translates into
\begin{equation}
\Gamma_{ij}(\vec{p})=\frac{1}{\beta}\delta_{ij}
-\frac{1}{\beta^2}\left\langle
\Delta^{\xy}_{i}(-\vec{p})\Delta^{\xy}_{j}(\vec{p})
\right\rangle_\xy.
\label{equ:dual2point}
\end{equation}
Note that
\begin{equation}
\chi_m=\lim_{\vec{p}\rightarrow 0}\Gamma_{33}(\vec{p}).
\label{equ:def-chim}
\end{equation}

More generally, we can see that there is 
a direct correspondence between the photon in the gauge theory and the
Noether current density $\Delta^{\xy}_{\vec{x},i}$ in the XY
model.

\section{Phase structure}
\subsection{Symmetric/superconducting phase}
\label{sect:symmphase}

\begin{table*}
\begin{tabular}{l|l|l}
XY: symmetric \hfill $\kappa<\kappa_c$ & 
critical  \hfill $\kappa=\kappa_c$ & 
XY: broken    \hfill $\kappa>\kappa_c$ \\
\fzs: superconducting \hfill $\beta>\beta_c$ & 
               \hfill $\beta=\beta_c$ & 
\fzs: Coulomb         \hfill $\beta<\beta_c$ \\
\hline
$\xi=\tens^{-1}
\sim|\kappa-\kappa_c|^{-\nu}$
& 
$G_{\vec{x},\vec{y}}\sim |\vec{x}-\vec{y}|^{-(1+\eta)}$
&
$\tens=0$\\
$\lambda=m_\gamma^{-1}\sim|\kappa-\kappa_c|^{-\nu'}$
&
$\Gamma_{ij}(\vec{p})\sim |\vec{p}|^{\eta_A}$
&$\lambda=m_\gamma^{-1}=\infty$\\
$\chi_A\sim|\beta-\beta_c|^{-\gamma_A}$
&&$\Upsilon\sim\chi_m\sim|\kappa-\kappa_c|^{\chimexp}$\\
\hline
\end{tabular}
\caption{
\label{tab:phases}
The behaviour of certain observables in the two phases of the models.}
\end{table*}

In the XY model, the symmetric phase ($\kappa<\kappa_c$)
is characterized by vanishing
magnetization $M$, and a finite correlation length $\xi=1/m$. 
The helicity modulus
$\Upsilon$ is zero, and therefore Eq.~(\ref{equ:upsilon-chi}) implies
that $\chi=1/\kappa$.
When the critical point is approached, 
the correlation length $\xi$ diverges as 
$\xi\sim|\kappa-\kappa_c|^{-\nu}$. Numerical studies [using the cosine
action rather than Eq.~(\ref{equ:Villain})] have shown that $\nu\approx
0.6723$.\cite{Hasenbusch:1999cc}

In the \fzs, this corresponds to the superconducting phase
($\beta>\beta_c$). The above implies that the vortex tension $\tens$,
which is equal to the scalar mass $m=1/\xi$ of the XY model,
is non-zero, and vanishes as $\tens\sim|\beta-\beta_c|^\nu$ at the
transition point. The magnetic permeability $\chi_m$, which is equal
to $\Upsilon$ vanishes. However,
we can define the gauge field susceptibility $\chi_A$ by
[cf.~Eq.~(\ref{equ:def-chim})]
\begin{equation}
\chi_A=\lim_{\twovec{p}\rightarrow
0}\frac{\Gamma_{33}(\twovec{p})}{\twovec{p}^2},
\label{equ:def-chiA}
\end{equation}
where the underline indicates that $\twovec{p}$ is a two-vector in the
$(1,2)$ plane. 
This quantity diverges as the critical point is approached, and we
parameterize this divergence by the exponent $\gamma_A$,
\begin{equation}
\chi_A\sim|\beta-\beta_c|^{-\gamma_A}.
\end{equation}
It was argued in Ref.~\onlinecite{olsson1998} that $\gamma_A=\nu$.

The photon correlator $\Gamma_{(\vec{x}_1,i)(\vec{x}_2,j)}$ decays
exponentially in this phase, and the penetration
depth
 $\lambda$ is
therefore finite. It diverges at the transition point as 
$\lambda\sim|\beta-\beta_c|^{\nu'}$. By duality, the current-current
correlation length of the XY model has the same behaviour.
There has been a lot of debate in the literature about the value of
$\nu'$. Originally, it was believed that
$\nu'=\nu/2$,\cite{fisher1991} but Kiometzis et 
al.\cite{kiometzis1994} later argued that
the penetration
depth
 does not get renormalized and would therefore
have the mean-field exponent $\nu'=1/2$. Finally, Herbut and 
Tesanovic\cite{herbut1996} found $\nu'=\nu$ using renormalization
group arguments, and this value was later confirmed in
Refs.~\onlinecite{olsson1998,calan1999,hove2000} with different 
approaches. However, all these results rely on some analytical
approximations. They also disagree with the results
of superconductor experiments.\cite{kamal1998,paget1999}

\subsection{Broken/Coulomb phase}
\label{sect:brokenphase}
{}From the XY model point of view, the characteristic property of the
broken phase ($\kappa>\kappa_c$) is non-zero magnetization $M\neq
0$. Another signal for symmetry breakdown is a non-zero value of the
helicity modulus $\Upsilon$. Through the duality, this corresponds to
non-zero magnetic permeability $\chi_m$ in the Coulomb phase of the
\fzs.

In the extreme high-temperature limit $\beta\rightarrow 0$, the \fzs\
approaches free non-compact electrodynamics, which is given by
Eq.~(\ref{equ:glpart}) in the limit $\kappa\rightarrow 0$. In this
case, it is easy to calculate $\chi_m$ from the path integral, and we
find
\begin{equation}
\chi_m\sim 1/\beta\quad\mbox{when $\beta\rightarrow 0$}.
\label{equ:lowkappa-chim}
\end{equation}
Using Eqs.~(\ref{equ:upsilon-chi}) and (\ref{equ:ups-equals-chim}),
we find $\Upsilon\rightarrow\kappa$ and 
$\chi\rightarrow 0$, as $\kappa\rightarrow\infty$.

As the critical point is approached $\chi_m$ vanishes, and 
following Ref.~\onlinecite{fisher1973}, we
parameterize this with the critical exponent
$\chimexp$, 
\begin{equation}
\chi_m\sim|\beta-\beta_c|^{\chimexp},\quad \beta\nearrow \beta_c. 
\end{equation}
By duality, the
helicity modulus $\Upsilon$ must behave in the same way. It has been
argued in Refs.~\onlinecite{fisher1973,olsson1998,son2002} that $\chimexp=\nu$.
The critical exponents are summarized in Table~\ref{tab:phases}.

\subsection{Critical point}
At the transition point, the spin-spin correlator in the XY model has
a power-law decay $G_{\vec{x}_1,\vec{x}_2}\sim
|\vec{x}_1-\vec{x}_2|^{-(1+\eta)},$ where the anomalous dimension
$\eta$ has been measured to be $\eta\approx
0.038$.\cite{Hasenbusch:1999cc}

Similarly, one can define the anomalous dimension $\eta_A$ for the photon
correlator in the \fzs\ by
\begin{equation}
\Gamma_{ij}(\vec{p})\sim |\vec{p}|^{\eta_A},\quad\mbox{when
$|\vec{p}|\rightarrow 0$}.
\end{equation}
Earlier studies have shown that $\eta_A\approx 
1$.\cite{Bergerhoff:1995zq,herbut1996,olsson1998,hove2000}

\begin{figure}[tb]
\centerline{ 
\psfig{file=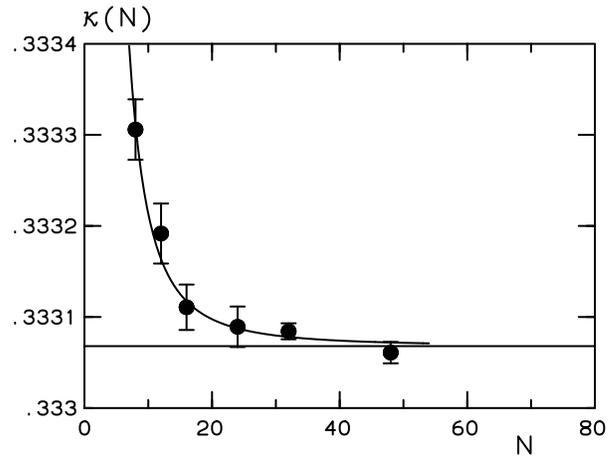,angle=270,width=8.5cm}
     }
\caption[a]{ Pseudocritical hopping parameter values $\kappa(N)$
             as a function of the system size $N$ in the XY model
             \peq{xy-partition-function}%
, together with a fit to the scaling ansatz in 
Eq.~(\ref{equ:pseudoscaling})%
. }
\label{fig:scalartheory-a}
\end{figure}

\begin{figure}
\psfig{file=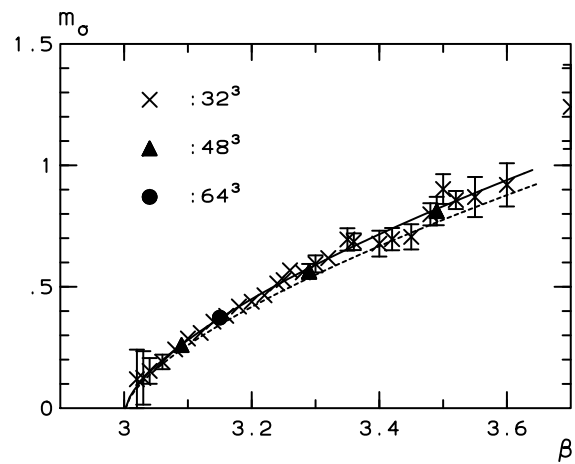,angle=270,width=8.5cm}
\caption[a]{The scalar
             mass $m$ in the XY model as a function of the
             dual coupling $\beta=1 / \kappa$. The solid curve shows
             the power-law fit in Eq.~(\ref{equ:mbeta}), and the
             dashed line the vortex tension fit in the \fzs~[see
             Eq.~(\ref{tension_scaling})].
           }
\label{fig:scalartheory-b}
\end{figure}

\section{Locating the critical point}

In order to explore the manifestations of the duality, we 
study both the \fzs\ \peq{iz-partition-function} and 
the XY model \peq{xy-partition-function} using Monte Carlo
simulations.  For the \fzs, the simulation algorithm consists 
of a single hit Metropolis update of the integer link variables, and for 
the XY model we use a Swendsen-Wang type reflection cluster 
algorithm.\cite{SwendsenWang}  One update sweep for the
\fzs\ consists of $N_1N_2N_3$ single link Metropolis hits, and for the
XY model out of one full Swendsen-Wang cluster update.  
The number of Monte Carlo update sweeps typically ranges from 
$50 000$ to $10^7$, depending on lattice sizes and coupling 
constant values. The errors are determined by 
jackknife error calculation.

Because of the duality, the critical points of the theories are
related by
$\beta_c = 1/\kappa_c$.  We determined the critical coupling values 
\begin{equation}
\kappa_c=0.333068(7), \quad \beta_c=3.00239(6)
\label{the-critical-point}
\end{equation}
by a numerical simulation of
the XY model on cubic $N=N_1=N_2=N_3$ lattices
with $N = 8$--96.  
For this, we used the Binder cumulant 
method:\cite{binder}
Matching the fourth-order cumulants of the magnetization
measured from lattices of size $N$ and $2N$ gives us an estimate of the
$N$-dependent pseudocritical hopping parameter $\kappa_c(N)$.  The
results are shown in Fig.~\ref{fig:scalartheory-a}.
Using the finite size scaling ansatz
\begin{equation}
\kappa_c(N) = \kappa_c + \frac{ A_\kappa}{N ^{{1/\nu}+\omega}},
\label{equ:pseudoscaling}
\end{equation}
with $\omega \approx 0.8$ denoting the subleading XY model 
scaling exponent,\cite{Hasenbusch:1999cc} we arrive at the numerical result 
in \eq{the-critical-point}.  
The uncertainty in this value is small enough to have a negligible effect
for the determination of the critical indices.

Most numerical studies of spin models with XY 
universality\cite{Hasenbusch:1999cc} 
use the cosine action $s(x)=-\cos(x)$ rather than Eq.~(\ref{equ:Villain}), and
subsequently there is only limited experience with the Villain action.
Nevertheless, there is no doubt that the phase transition with the 
Villain action 
is of second order and in the same universality class as with the 
cosine action.
As an illustration we show the scalar mass $m=1/\xi$ in the symmetric 
phase ($\kappa < \kappa_c$), as a 
function of the ``temperature'' $\beta=1 / \kappa$,
in Fig. \ref{fig:scalartheory-b}.
The solid curve in the figure shows a $\chi^2$-fit 
with the scaling law
\begin{equation}
m(\beta) =  A_{\xi} (\beta-\beta_c)^{\nu}
\label{equ:mbeta}
\end{equation}
and with a fit with $\chi^2/\mbox{d.o.f.}=0.35$
we obtain the parameter values $A_{\xi}=1.32(4)$
and $\nu=0.66(2)$, in agreement with the cosine action 
results.\cite{Hasenbusch:1999cc}
The amplitude value $A_{\xi}$ 
is important for the comparison with the
vortex tension in the gauge theory, to be discussed below.

\begin{figure}[tb]
\centerline{ 
   \psfig{file=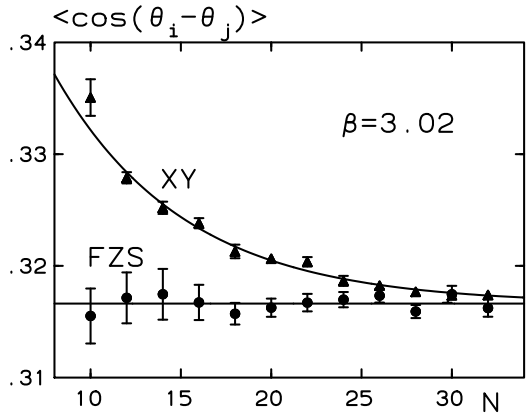,angle=270,width=8.8cm} 
    }
\caption[a]{ A numerical check of duality relations for the
             nearest-neighbour correlator in
             Eq.~(\ref{equ:nearestneighbour}) on a finite lattice
as a function of the linear lattice size $N$%
.
           }
\label{fig:check-duality}
\end{figure}

\section{Vortex tension}
\label{sec:tension}

In order to illustrate the numerical consequences of the duality
we measured the nearest-neighbour spin-spin correlation function,
\begin{equation}
G_{\vec{x}+i,\vec{x}}
=\langle {\rm cos}( \theta_{\vec{x}+i}-\theta_{\vec{x}} )
\rangle_{\xy}
\label{equ:nearestneighbour}
\end{equation}
in the XY model at $\beta=1/\kappa=3.02$. In the 
gauge theory, this is mapped to the expectation value of 
\begin{equation}
\langle e^{ \beta \plaq_{\vec{x},ij}- \beta/2 }\rangle_{\fzs}.
\end{equation}
In Fig.~\ref{fig:check-duality} we compare both measurements as
functions of the system size $N$ on cubic boxes with periodic
boundary conditions.
It is evident that Eq.~(\ref{equ:hplaq}), and thereby the duality, 
only is valid for infinite
systems, i.e., the boundary conditions of finite systems do not respect
our duality arguments. The curves in the figure assume finite volume corrections
of the form ${\rm exp}(-cN)$ and extrapolate to the same value in the 
thermodynamic limit.

Let us now consider the long-distance behaviour of the 
correlation function $G_{\vec{x}_1,\vec{x}_2}$
defined in Eq.~(\ref{equ:xycorr}).
The duality maps this correlator into 
Eq.~(\ref{equ:dualcorr}) in the \fzs.
In principle, we could measure Eq.~(\ref{equ:dualcorr}) directly, but
it turns out to be more convenient to use an indirect approach, 
which was introduced in
Ref.~\onlinecite{Kajantie:1999zn}. 

As we shall now show, Eq.~(\ref{equ:dualcorr}) 
corresponds exactly to a vortex, or an Abrikosov flux tube, 
in the \fzs.   Thus, the
spin-spin correlation length $\xi$ in the XY model is exactly the tension $\tens$, 
the vortex free energy per unit length, in the \fzs.
These flux tubes exist because a superconductor resists 
applied external magnetic fields.
For field strengths above the critical field $H_{c1}$
magnetic flux penetrates the material, and the flux arranges 
into vortex lines, Abrikosov flux tubes.

In practice, we define the vortex tension as
follows: Using appropriate modifications of periodic boundary
conditions, to be described below, we constrain the net number
of vortex lines $n_V$ winding around the finite volume to, say, $z$ 
direction.  The vortex tension $\tens$ is then defined by
\begin{equation}
\tens = - \frac{1}{N_3} {\rm ln} [\frac{ Z_{\fzs}(n_V=1)}{Z_{\fzs}(n_V=0) }],
\label{equ:tension}
\end{equation}
in the $N_3 \to \infty$ limit.

Let us now discuss how the appropriate boundary conditions are constructed.
Labelling the cartesian coordinates $\vec{x}=(n_1,n_2,n_3)$ , $n_i=1,...,N_i$
$i=1,...,3$, and using periodic boundary conditions to all directions
\begin{eqnarray}
I_{ (n_1=N_1+1,n_2,n_3),i}=I_{ (n_1=1,n_2,n_3),i }, \nonumber \\ 
I_{ (n_1,n_2=N_2+1,n_3),i}=I_{ (n_1,n_2=1,n_3),i }, \\ 
I_{ (n_1,n_2,n_3=N_3+1),i}=I_{ (n_1,n_2,n_3=1),i }, \nonumber
\end{eqnarray}
the total magnetic flux $\Phi = 2\pi n_V$ through any planar 
2-dimensional cross section $\Omega_{i,j}$ of the lattice
\begin{equation}
\Phi(\Omega_{i,j}) = 
2\pi \sum_{p \in \Omega_{i,j}} I_p(i,j),~~~i \ne j
\end{equation}
equals zero because of the periodicity.  However, if we choose
a set of links $I_{(n_1,n_2,n_3),2}$, with $n_1=1$, $n_2=1$ and $n_3 = 1\ldots N_3$,
and use a ``twisted'' boundary condition
\begin{equation}
 I_{ (N_1+1,1,n_3),2}
=I_{ (1,1,n_3),2}+n_V,~~~n_3=1,...,N_3~,
\label{specialboundary}
\end{equation}
the flux through $N_3$ cross sectional areas $\Omega_{1,2}$
acquires the value $\Phi(\Omega_{1,2})=2\pi n_V$. At $n_V=1$ this  
corresponds to a single vortex of length $L=N_3$. The vortex line forms a
closed loop through the $z$-direction of the box.  

It should be noted that the modification of the boundary condition
does not
give the modified link a special status: by a suitable redefinition of the
link variables $I_j$, the ``twist'' can be moved to arbitrary location along
the (1,2)-plane, also away from the boundary.

\begin{figure}[tb]
\centerline{ 
\psfig{file=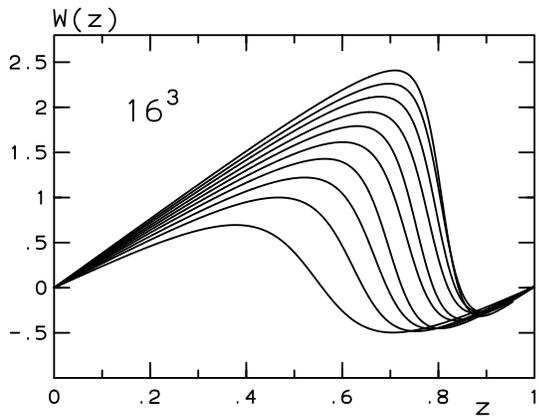,angle=270,width=8cm}
}
\caption[a]{ The function
             $W(z)$ 
[see Eqs.~(\ref{integral}) and (\ref{w_of_z})]
             measured on a  $16^3$ lattice, with $\beta$ varying
             from 3.0 (bottom curve) to 4.0 (top). }
\label{fig:tension-data-a}
\end{figure}

In practice, a more convenient way to implement the fixed flux is to
consider a system with fully periodic boundary conditions but with a modified
action: Let us define a plaquette field which is non-zero only on one
fixed ``stack'' of (1,2)-plaquettes,
\begin{equation}
  m_{\vec{x},i} = \left\{ \begin{array}{l}
                            1 \mbox{~~~ if $x_1 = x_2 = 1$; $i = 3$}, \\
                            0 \mbox{~~~ otherwise}
                            \end{array} \right.\,.
\end{equation}
Now we can define the partition function
\begin{eqnarray}
&&Z_{\fzs}(\beta,n_V)=\nonumber\\
&&~~\sum_{\{I_{\vec{x},i}\}} \exp\left[- \frac{\beta}{2} \sum_{\vec{x},i}
\left(\fr12 \epsilon_{ijk}  \plaq_{\vec{x},jk}- n_V
m_{\vec{x},i}\right)^2
\right], 
\label{iz-partition-function-with-twist}
\end{eqnarray}
which is equivalent to the partition function with the unmodified 
action but with the boundary condition \eq{specialboundary}. (Again,
the choice $x_1 = x_2 = 1$ here is arbitrary.)

Comparison of \eq{iz-partition-function-with-twist} with
\eq{equ:dualcorr} shows that the vortex tension in \eq{equ:tension} is
closely 
related to the XY model spin correlator $G_{\vec{x}_1,\vec{x}_2}$,
provided that we identify $m_{\vec{x},i}$ with $\alpha_{\vec{x},i}$.
Indeed, the only difference is that $m_{\vec{x},i}$ does not have start or end
points, but stretches across the whole system. 
However, on a periodic lattice,
we can imagine moving the source $\vec{x}_1$ of
$\alpha_{\vec{x},i}$ through the boundary to $\vec{x}_2$ so that it
cancels the sink. This leads to a sourceless field which has one
field line passing through the lattice,
and this shows that
in the infinite volume limit the XY model correlation length $\xi$
and the vortex tension $\tens$ are related by
\begin{equation}
\tens=1/\xi=m.
\label{equ:tensionmass}
\end{equation}

As an aside, we note that this procedure does not make sense for the
XY model spin correlators, since now $G_{\vec{x},\vec{x}+\hat{e}_3N_3}
= G_{\vec{x},\vec{x}}=1$, again showing that the periodic boundary
conditions in a finite volume do not respect the duality relation.

In practice, all measurements are affected by the finite size of the
lattices and by the performance of the numerical algorithms, and it is 
therefore interesting to compare
the direct measurement of $\tens$ with the measurement of $m$ in the
XY model.
In order to measure $\tens$ in simulations, we write it
as an integral
\begin{equation}
\tens =  \int_0^1~W(z)~dz
\label{integral}
\end{equation}
over an expectation value $W(z)$
\begin{widetext}
\begin{eqnarray}
W(z) &\equiv& \frac{\partial \log Z_{\fzs}(\beta,z)}{\partial z}
\nonumber\\
&=& \frac{1}{Z_{\fzs}(\beta,z)} \frac{\beta}{N_3}
\sum_{\{I_{\vec{x},i}\}}
\left[
\sum_{\vec{x},i}
	(\fr12 \epsilon_{ijk} \plaq_{\vec{x},jk}-zm_{\vec{x},i})m_{\vec{x},i}
\right] 
\exp\left(- \frac{\beta}{2}  \sum_{\vec{x},i}
(\fr12 \epsilon_{ijk} \plaq_{\vec{x},jk}-zm_{\vec{x},i})^2
\right) ,
\label{w_of_z}
\end{eqnarray}
\end{widetext}
which can determined with Multicanonical Monte Carlo simulations
(for technical details, we refer to Ref.~\onlinecite{Kajantie:1999zn}).
In Fig.~\ref{fig:tension-data-a}, we show
the function $W(z)$ measured from a $16^3$ volume as 
a function of $z$ for several values of $\beta$.
$\beta$-values in the superconducting phase and close
to the critical point.

\begin{figure}[tb]
\centerline{
\psfig{file=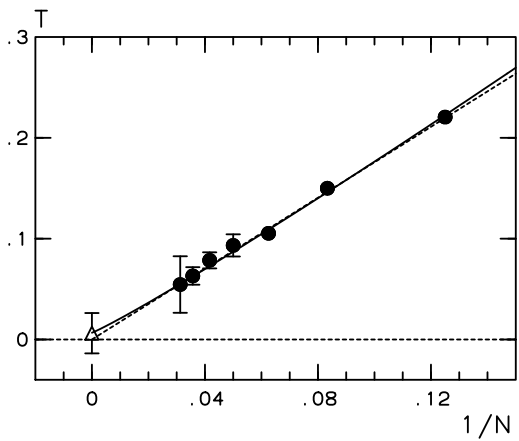,angle=270,width=8cm}
}
\caption[a]{ 
 	The tension $\tens$ 
	     at $\beta=3.0016$ (slightly below $\beta_c$, i.e.
             in the Coulomb phase)
             as a function of $N^{-1}$. In the infinite-volume limit,
             the tension 
             extrapolates
             to zero.
The relationship $\tens\propto 1/N$ is equivalent to 
$B\propto H^2$ predicted in
Ref.~\onlinecite{son2002}. 
           }
\label{fig:tension-data-b}
\end{figure}

\begin{figure}[tb]
\centerline{ 
   \psfig{file=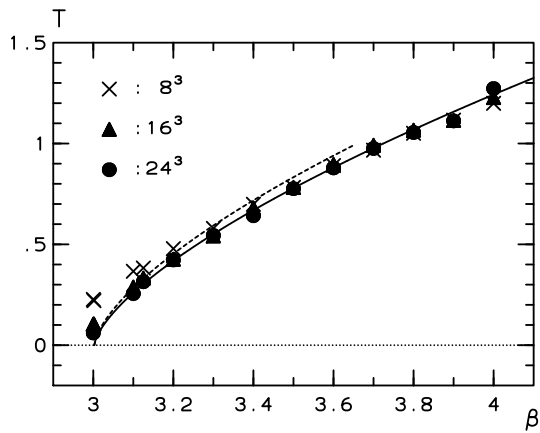,angle=270,width=8cm} 
 }
\caption[a]{ The 
vortex
tension $\tens$ on $8^3$, $16^3$ and $24^3$
             lattices and a fit to $24^3$ data
             in the superconducting phase of
             the \fzs. The solid line corresponds to the fit in
             Eq.~(\ref{tension_scaling}) and the dashed line to the XY
             model mass in Eq.~(\ref{equ:mbeta}).
           }
\label{fig:the-tension}
\end{figure}

The vortex tension $\tens$ vanishes in 
the Coulomb phase and at criticality $\beta=\beta_c$. 
In Fig.~\ref{fig:tension-data-b} we show the measured value of $\tens$
as a function of the lattice size $N$, using $\beta=3.0016$, which is very close to
$\beta_c$ in the Coulomb phase.
The tension decreases with increasing linear size $N$, and assuming finite size
corrections of the form $N^{-\alpha}$ we obtain the fit
\begin{equation}
\tens(N)= 0.006(26) + {2.0(7)~N^{-\alpha} },
 ~~~~~~~ \alpha=1.1(2),
\label{tension_scaling_at_criticality}
\end{equation}
with $\chi^2/\mbox{d.o.f.} = 0.50$ for the fit.
The data support a vanishing tension in the Coulomb phase
close to criticality. The value of $\alpha$ is consistent
with $\alpha=1$ and, at criticality and in the dual XY model
$\alpha=1$ corresponds to a scalar correlation length finite size 
scaling $\xi(N) \propto N$.  
Fixing $\alpha=1$ we obtain the fit
\begin{equation}
\tens(N)=  { 1.76(3) \over N }  ~~~~~ \beta \approx \beta_c
\label{tension_scaling_at_criticality_2}
\end{equation}
at the $\chi^2/\mbox{d.o.f.}$ value of $0.65$ for the fit.

In fact, the relation $\tens\propto 1/N$ is equivalent to 
a conjecture by Son\cite{son2002} that the flux density $B$
should be proportional to the square of the external field $H$.
In continuum normalization, the flux density corresponding to one
vortex is
\begin{equation}
B=\frac{2\pi}{e}\frac{1}{\delta x^2N^2},
\label{equ:BN}
\end{equation}
and the external field needed to create it is given by
\begin{equation}
H=\frac{e}{2\pi}\frac{\tens}{\delta x},
\label{equ:HT}
\end{equation}
where the coupling constant $e$ and the lattice spacing $\delta x$ are
dimensionful quantities.
Son defined the constant of proportionality $C$ by
\begin{equation}
B=\left(\frac{2\pi}{e}\right)^3CH^2,
\end{equation}
and argued that it should be a universal quantity. Using
Eqs.~(\ref{equ:BN}) and (\ref{equ:HT}), we can rephrase this as
\begin{equation}
\tens=C^{-1/2} N^{-1},
\end{equation}
and Eq.~(\ref{tension_scaling_at_criticality_2}) tells us that $C=0.32(1)$.


In the superconducting phase, we calculated the vortex tension $\tens$ 
using $8^3$, $16^3$ and $24^3$ lattices; the
results are presented in Fig.~\ref{fig:the-tension} as functions of $\beta$.
The finite size effects between $16^3$ and $24^3$ lattices 
are smaller than the statistical errors. 
We fit the $24^3$ data in a broad scaling 
interval $3.05< \beta <  4.1$ with the power-law behaviour
\begin{equation}
\tens(\beta) =  A_{\tens} (\beta-\beta_c)^{\tensexp},
\label{tension_scaling}
\end{equation}
where $\tensexp$ denotes the tension scaling exponent.
We obtain the parameter values $A_{\tens}=1.24(2)$
and $\tensexp=0.672(9)$ with a $\chi^2/\mbox{d.o.f.}$ value
of $1.13$ for the fit. 

The duality (\ref{equ:tensionmass}) implies that $A_{\tens}$ and $\tensexp$
ought to be equal to the XY model quantities $A_\xi$ and $\nu$
in \eq{equ:mbeta}.  Indeed, 
we find that $\tensexp$ agrees with the XY model exponent
$\nu=0.671$, and that
$A_{\tens}/A_{\xi} \approx 0.94(5)$.
In summary,
the vortex tension $\tens(\beta)$ of the gauge theory 
agrees within two standard deviations
of statistical errors with the scalar mass
$m(\kappa=1 / \beta)$ of the XY model.
This constitutes a highly non-trivial test for the methods developed
in Ref.~\onlinecite{Kajantie:1999zn}.


\section{SUSCEPTIBILITIES}
\label{sect:susceptibilities}

We shall now discuss the susceptibilities $\chi_m$ and $\chi_A$
defined in Eqs.~(\ref{equ:def-chim}) and (\ref{equ:def-chiA}),
respectively. We also generalize the definition of $\chi_A$ 
to non-zero momenta by
\begin{equation}
\chi_A(\vec{p})=
\frac{\Gamma_{33}(\vec{p})}{p_{\Lambda(1,2)}^2},
\label{equ:chiAp}
\end{equation}
where $p_{\Lambda(1,2)}^2$ is defined as
\begin{equation}
{p^2_{\Lambda(1,2)}} = 2 \sum_{i=1}^2 [1-\cos p_i]
= 2 \sum_{i=1}^2 [1-{\rm cos}(2 \pi k_i /N_i)] .
\end{equation} 
We use the lowest
momentum value for the $(1,2)$-plane momentum, but 
non-zero values of the $z$-component of the momentum 
$p_3=2\pi k_3/N_3$.
We expect the momentum dependence of the susceptibility 
to be a function of the
total lattice momentum squared
\begin{equation}
{p^2_{\Lambda}} = 2 \sum_{i=1}^3 [1-\cos p_i]
= 2 \sum_{i=1}^3 [1-{\rm cos}(2 \pi k_i /N_i)] .
\end{equation} 
As discussed in Sections~\ref{sect:symmphase} and
\ref{sect:brokenphase}, 
at zero momentum 
$\chi_A$ is finite and non-zero in the superconducting phase and
diverges in the Coulomb phase, whereas
$\chi_m$ vanishes in the
superconducting phase and is non-zero in the Coulomb phase, being equal
to the helicity modulus $\Upsilon$ of the XY theory.

\begin{figure}[tb]
\centerline{ 
 \psfig{file=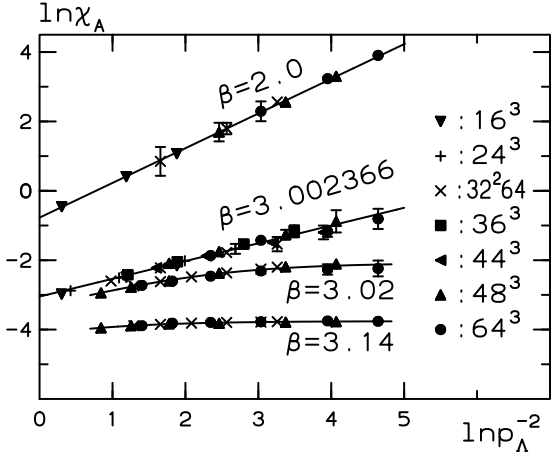,angle=270,width=8.5cm}
}
\caption[a]{
 	Logarithm of the gauge field susceptibility 
             $\ln \chi_A(\vec{p})$
[see Eq.~(\ref{equ:chiAp})]
             as a function of $\ln {{p}}_{\Lambda}^{-2}$ in the Coulomb 
             phase at $\beta=2.0$ (top curve), at
             the critical point $\beta=\beta_c$ (second curve from above), 
             and for
             two $\beta$-values in the superconducting phase.
Different symbols correspond to different lattice sizes.
}
\label{fig:singularity}
\end{figure}

\begin{figure}[tb]
\centerline{ 
 \psfig{file=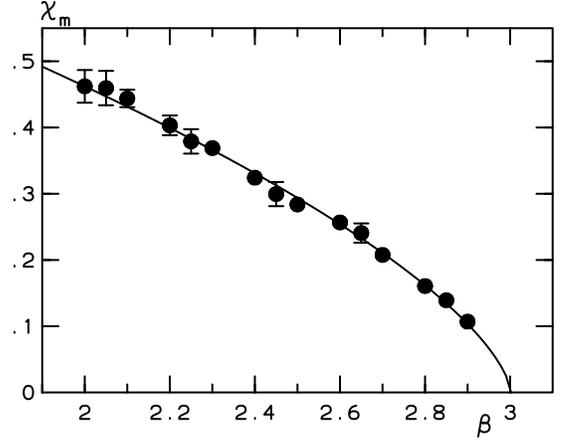,angle=270,width=8.5cm}
}
\caption[a]{
             The magnetic permeability $\chi_m$
[see Eq.~(\ref{equ:ups-equals-chim})]
             as a function of $\beta$
             in the Coulomb phase. The curve corresponds to the
             fit in \eq{z_val_scaling}.
           }
\label{fig:singularity-b}
\end{figure}

Our measurements demonstrate the presence of a 
massless photon pole in the Coulomb phase of the gauge theory.
At $\beta=2.0$ -- well within the Coulomb phase -- we 
determine the susceptibility $\chi_A(\vec{p})$ on $16^3$, $32^3$, $48^3$ 
and $64^3$ lattices as a function of small momentum values 
$\vec{p}$.  The results are shown in Fig.~\ref{fig:singularity} 
as a function of ${\rm ln}~{{p}}_{\Lambda}^{-2}$.
The data are fitted with the form
\begin{equation}
\chi_A(\vec{p}) = \frac{ 0.46(1)}{{p}^2_{\Lambda}}~~~~\beta=2.0,
\end{equation}
with $\chi^2_{d.o.f.} = 0.67$ for the fit. This is equivalent
to saying that in the zero-momentum limit, $\chi_m=0.46(1)$ 
at $\beta=2.0$. This value is slightly smaller than the result 
from free electrodynamics in Eq.~(\ref{equ:lowkappa-chim}).  

In fact, since $\chi_A(\vec{p})$ is essentially the gauge field
correlation function, albeit in a gauge-invariant form, the divergence
$1/p_\Lambda^2$ demonstrates the presence of a massless
photon. Thus, we can interpret $\chi_m=\Upsilon$ as the corresponding
wave function renormalization $Z$ factor.

As the critical point is approached 
from the Coulomb phase $\chi_m$ vanishes with the exponent
$\chimexp$, as discussed in Section~\ref{sect:brokenphase}.
The measured values of $\chi_m$ are presented in Fig.~\ref{fig:singularity-b},
together with the fit to the scaling ansatz 
\begin{equation}
\chi_m(\beta) =  A_{\Upsilon} (\beta_c-\beta)^{\chimexp}.
\label{z_val_scaling}
\end{equation}
The parameter values are $A_{\Upsilon}=0.46(1)$ and 
$\chimexp=0.66(2)$ and the fit has a $\chi^2_{d.o.f.}$ value of $0.32$. 
By duality, this is exactly the scaling of $\Upsilon$ near the
critical point in the XY model, and the value of $\chimexp$ is indeed
entirely consistent with arguments that $\Upsilon$ scales
with the exponent $\nu$.\cite{fisher1973} The amplitude
value $A_{\Upsilon}$ could also be compared directly with the XY model, but
we are not aware of helicity modulus data for the
Villain action.

\begin{figure}[tb]
\centerline{ 
   \psfig{file=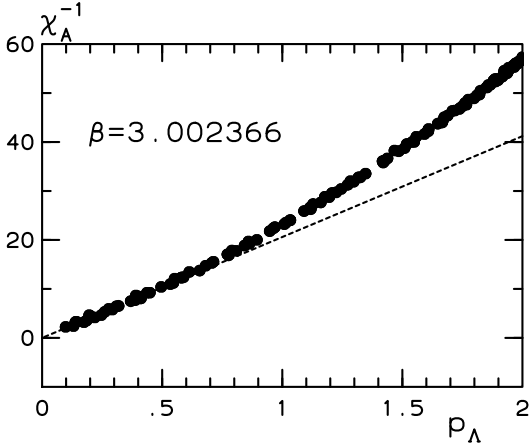,angle=270,width=9cm} 
    }
\caption[a]{ The inverse gauge susceptibility  $\chi_A^{-1}(\vec{p})$ at
             the critical point $\beta=3.002366$ 
             for $16^3,24^3,32^2 \times 64,36^3,44^3,48^3$
             and $64^3$ lattices as a function of
             $p_{\Lambda}$. 
             The dashed line
             corresponds to \eq{small_momentum_behavior}.
           }
\label{fig:critical-dispersion}
\end{figure}

The anomalous dimension $\eta_A$ of the gauge field at criticality
can be obtained from the momentum dependence of the gauge
field susceptibility $\chi_A(\vec{p})$ as
\begin{equation}
\chi_A^{-1}(\vec{p}) = c_A({p}_{\Lambda}^2)^{1-\eta_A/2}
~~~~~\beta=\beta_c,
\label{pole_structure}
\end{equation}
for low $\vec{p}$.  We show the data together with a power law fit
in 
Fig.~\ref{fig:singularity},
as well as in Fig.~(\ref{fig:critical-dispersion}).
The fit to Eq.~(\ref{pole_structure}) gives
the exponent
\begin{equation}
\eta_A=0.98(4),
\label{eta_value}
\end{equation}
which is perfectly consistent with the value $\eta_A=1$
predicted in Refs.~\onlinecite{Bergerhoff:1995zq,herbut1996}. 
If we fix $\eta_A=1$, we obtain the critical amplitude
\begin{equation}
c_A=20.7(3), 
\label{critical-amplitude}
\end{equation}
which implies
\begin{equation}
\chi_A(\vec{p})  =   {0.0484(7) \over  p_{\Lambda}},
\label{small_momentum_behavior}
\end{equation}
for small momenta.
This fit is shown as a straight dashed line in Fig.~\ref{fig:critical-dispersion}.

In the superconducting phase of the \fzs\  
the magnetic permeability $\chi_m$ vanishes at zero
momentum, and therefore the photon becomes massive.
The gauge field susceptibility $\chi_A(\vec{p})$ tends to a finite limit
as $\vec{p}\rightarrow 0$.  This is shown in Fig.~\ref{fig:singularity}, 
where the lowest two data sets of the figure correspond to $\chi_A$ in 
the superconducting phase. 

%
%
At the critical point ($\beta=\beta_c$), 
$\chi_A(\vec{p})$ is proportional to $p^{\eta_A/2-1}$ 
[see \eq{pole_structure}].
The phase transition in the {\fzs} is continuous, and therefore 
the analytic form of $\chi_A(\vec{p})$ at non-zero momenta must 
interpolate smoothly 
between the critical, massless behaviour and the massive mode
in the superconducting phase.
This reasoning leads to the 
Fisher scaling relation
\begin{equation}
\nu' = { \gamma_A \over 2-\eta_A }.
\label{fishers_scaling}
\end{equation}
Perhaps the simplest way in which this could happen is if the
susceptibility has the form
\begin{equation}
\chi_A^{-1}(\vec{p}) =
c_A  \left( p_{\Lambda}^{2} + m_\gamma^{2} \right)^{1-\eta_A/2},
\label{ansatz}
\end{equation}
which would correspond to
the asymptotic behaviour
\begin{equation}
\Gamma(\tau)\equiv
\Gamma(\tau,\twovec{p}=0)\sim \tau^{-\eta_A/2}\exp(-m_\gamma \tau),
\label{equ:besselKasymp}
\end{equation}
of the photon correlation function.  

In the $\vec p = 0$ limit Eq.~(\ref{ansatz}) 
relates the photon mass and the gauge field susceptibility $\chi_A$ through
\begin{equation}
\chi_A=\frac{1}{c_A m_\gamma^{2-\eta_A}}.
\label{equ:chiAmgamma}
\end{equation}

\begin{figure}[tb]
\centerline{ 
   \psfig{file=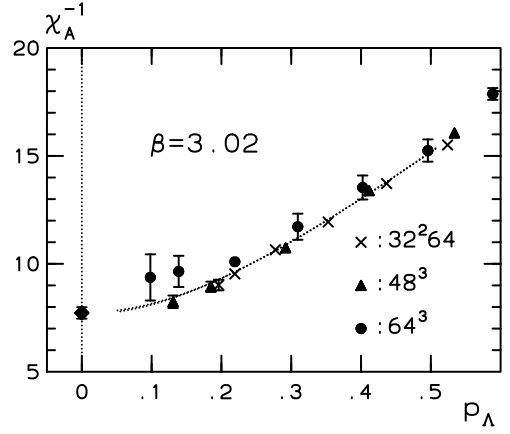,angle=270,width=8.5cm} 
   }
\caption[a]{ $\chi_A^{-1}(\vec{p})$ at
             $\beta=3.02$ in the superconducting phase 
             for $32^2 \times 64,48^3$
             and $64^3$ lattices as a function of
             $p_{\Lambda}$. 
             The two dotted lines, which are barely distinguishable, 
	     correspond to fits 
             using Eqs.~(\ref{ansatz}) and (\ref{equ:arctan}) and
             extrapolate 
             to finite values at zero momentum.
           }
\label{fig:non-critical-dispersion}
\end{figure}

While the precise form of the ansatz does not affect the
scaling relation (\ref{fishers_scaling}), it would lead to a
systematic error in the determination of $m_\gamma$. 
To estimate these errors, we also consider an ansatz based on treating
the dual XY model as a theory of a free complex scalar field (see
Appendix~\ref{sect:app}),
\begin{equation}
\chi_A(\vec{p})=\frac{4}{\pi}\frac{1}{c_Ap^2}\left[
\frac{m_\gamma}{2} + {p^2-m_\gamma^2\over 2p}\arctan {p\over m_\gamma}\right].
\label{equ:arctan}
\end{equation}
Again, $m_\gamma$ gives the exponential decay rate of the correlator
and at the critical point, the ansatz agrees with
Eq.~(\ref{pole_structure}). However, Eq.~(\ref{equ:chiAmgamma})
becomes
\begin{equation}
\chi_A=\frac{8}{3\pi}\frac{1}{c_A m_\gamma}.
\label{equ:chiAmgamma2}
\end{equation}
Therefore, we can estimate that the systematic error in the
determination of $m_\gamma$ is roughly $3\pi/8-1\approx 20\%$.

As an example, we show $\chi_A^{-1}$ at $\beta=3.02$ in
Fig.~\ref{fig:non-critical-dispersion}, together with fits of momenta
$0<p_\Lambda<0.5$ to Eqs.~(\ref{ansatz}) and (\ref{equ:arctan}).  
There is practically no
difference between the fits, 
but the fit parameters $c_A$ and $m_\gamma$
are different.

\begin{figure}[tb]
\centerline{ 
 \psfig{file=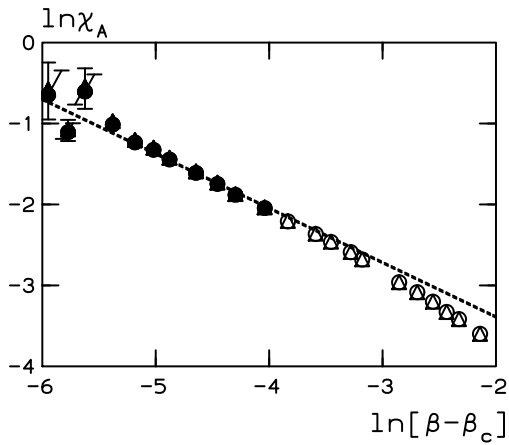,angle=270,width=8.5cm}
}
\caption[a]{  $\ln \chi_A$
             as a function of  $\ln (\beta-\beta_c)$ in the  
             superconducting phase, together with power law fits
to Eq.~(\ref{equ:chiApower}). 
             Here, as well as in the other figures, the data points that
             were included in the fit are indicated by filled symbols 
             and the others by open symbols.
}
\label{fig:final-result-a}
\end{figure}

\begin{figure}
\centerline{
 \psfig{file=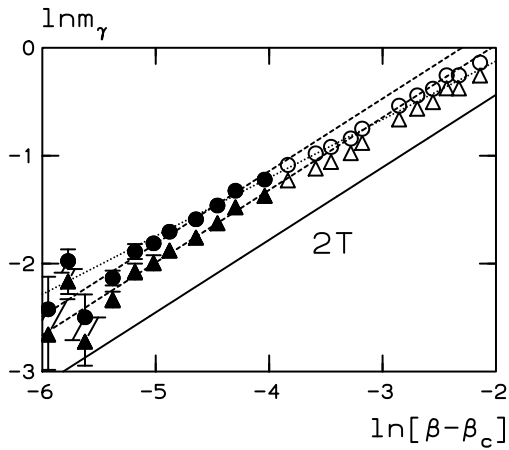,angle=270,width=8.5cm}
 }
\caption[a]{ The
             logarithm of the photon (pole) mass $m_\gamma$, together
             with power-law fits (dashed lines) and the theoretical
             expectation $m_\gamma=2\tens$.             
             The
             circles and triangles correspond to zero-momentum
             extrapolations using ans\"atze (\ref{ansatz}) and
             (\ref{equ:arctan}), respectively. The dashed lines show
             fits to the filled symbols with the XY model exponent
             $\nu'=\nu_{\rm XY}$. The dotted line is the same as the
             solid line in Fig.~\ref{fig:photon-mass-b}.
           }
\label{fig:final-result-b}
\end{figure}

Performing the fit to Eqs.~(\ref{ansatz}) and (\ref{equ:arctan})
at several $\beta >
\beta_c$ we obtain $\chi_A \equiv \chi_A(\vec p \rightarrow 0)$ and $m_\gamma$ 
as functions of $\beta$.  The results for $\chi_A(\beta)$ and
$m_\gamma(\beta)$ are shown in Figs.~\ref{fig:final-result-a} and \ref{fig:final-result-b}.  

The gauge field susceptibility $\chi_A(\beta)$ clearly has a
power law divergence as we approach the critical point.  Fitting the ansatz
\begin{equation}
\chi_A = A_{\chi} (\beta-\beta_c)^{-\gamma_A}
\label{equ:chiApower}
\end{equation}
to the datapoints
shown by full symbols in Fig.~\ref{fig:final-result-a},
we obtain
\begin{eqnarray}
 \gamma_A&=&0.68(3)\quad
\mbox{for Eq.~(\ref{ansatz}),}\nonumber\\
 \gamma_A&=&0.70(4)\quad
\mbox{for Eq.~(\ref{equ:arctan}),}
\label{equ:gammaA}
\end{eqnarray}
both of which are compatible with the result $\gamma_A=\nu_{\rm XY}$
obtained in
Ref.~\onlinecite{olsson1998}. 
Assuming $\gamma_A=\nu_{\rm XY}$, we obtain
\begin{eqnarray}
 A_{\chi}&=&0.00871(15)\quad
\mbox{for Eq.~(\ref{ansatz}),}\nonumber\\
 A_{\chi}&=&0.00888(17)\quad
\mbox{for Eq.~(\ref{equ:arctan}).}
\label{equ:fitAchi}
\end{eqnarray}

The critical behaviour of the photon mass $m_\gamma(\beta)$ 
(or equivalently, the penetration 
depth)
is more subtle.  
Inserting the measured values for $\eta_A$ and $\gamma_A$
in \eq{fishers_scaling}, we obtain
the critical exponent 
\begin{eqnarray}
\nu'&=&0.67(4)\quad
\mbox{for Eq.~(\ref{ansatz}),}\nonumber\\
\nu'&=&0.69(5)\quad
\mbox{for Eq.~(\ref{equ:arctan}).}
\end{eqnarray}
This result is fully compatible with the prediction 
$\nu'=\nu_{\rm XY}$.\cite{herbut1996,olsson1998,calan1999,hove2000}

\begin{figure}[tb]
\centerline{ 
\psfig{file=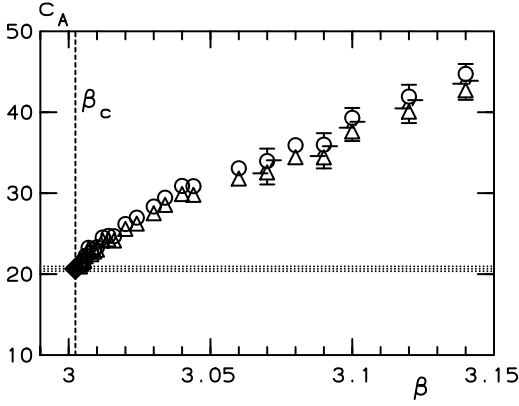,angle=270,width=8.5cm}
  }
\caption[a]{
The value of 
the constant
$c_A$ determined from fits to Eq.~(\ref{ansatz})
[circles] and Eq.~(\ref{equ:arctan}) [triangles].
           }
\label{fig:cA}
\end{figure}

Similarly, the scaling ans\"atze fix the critical
scaling function $m_\gamma = A_\gamma (\beta-\beta_c)^{\nu'}$
through Eqs.~(\ref{equ:chiAmgamma}) and (\ref{equ:chiAmgamma2}),
\begin{eqnarray}
A_{\gamma} &=& {1  \over c_A(\beta_c) A_{\chi} }
          = 4.69(8), \quad \mbox{for Eq.~(\ref{ansatz}),}\nonumber\\
A_{\gamma} &=& \frac{8}{3\pi}{1  \over c_A(\beta_c) A_{\chi} }
          = 3.94(6), \quad
\mbox{for Eq.~(\ref{equ:arctan}).}
\label{eq:ampltidude_relation}
\end{eqnarray}
The corresponding critical scaling functions for $m_\gamma$ are shown
in Fig.~\ref{fig:final-result-b} as dashed lines.  The scaling
functions only agree with direct mass measurements in a
narrow region near the critical point.  In fact, the direct mass
measurements would favour slightly critical exponents for the photon
mass
\begin{eqnarray}
\nu'&=& 0.58(3), \quad \mbox{for Eq.~(\ref{ansatz}),}\nonumber\\
\nu'&=& 0.62(3), \quad \mbox{for Eq.~(\ref{equ:arctan}).}
\label{eq:nuprimedirect}
\end{eqnarray}
The reason for this discrepancy is that $c_A$ changes rather rapidly
as 
a function of $\beta$ when the critical point is approached. This is
shown in Fig.~(\ref{fig:cA}). Using the critical value of $c_A$ in 
Eqs.~(\ref{equ:chiAmgamma})
and (\ref{equ:chiAmgamma2}) is therefore justified only very close to
the critical point.

Since the photon mass and the vortex tension scale with the same
exponent, it makes sense to calculate the 
the amplitude ratio  ${A_{\gamma}/ A_{\tens}}$.
We find
\begin{eqnarray}
A_\gamma/ A_\tens&=& 3.8, \quad \mbox{for Eq.~(\ref{ansatz}),}\nonumber\\
A_\gamma/ A_\tens&=& 3.2, \quad \mbox{for Eq.~(\ref{equ:arctan}).}
\end{eqnarray}
These values are significantly greater than the expected value of 2,
which follows from the assumption that if $m_\gamma>2\tens$,
the photon should be able to decay into two vortices,\cite{Peskin:1978kp}
and this would
result in exponential decay with rate $2\tens$.
We see two plausible reasons for this behaviour: Firstly,
it is likely that our ans\"atze
(\ref{ansatz}) and (\ref{equ:arctan}) are not of the right form.  In
this case, it would be important to find a theoretically better
motivated ansatz, which could then be tested numerically by
fitting it to our measurements. The correct functional form should
yield $A_\gamma/A_\tens=2$.  Secondly, the volumes available for
our analysis may be too small.  Vortex-vortex
interactions are presumably not negligible if the lattice size is of order
$1/m_\gamma$, which modifies the ``free vortex'' behaviour 
$m_\gamma = 2\tens$.

On the other hand, it may also be possible that the system has a
massive photon state, similar to resonances in particle physics, which
would eventually decay into two vortices at distance longer than what
we can probe in our simulations.



\section{PHOTON MASS}
\label{sec:photonmass}

In this section, we shall discuss the determination of 
the penetration
depth
 $\lambda$, or
equivalently the  photon mass $m_\gamma=1/\lambda$, 
directly from the decay of the correlator $\Gamma(\tau,\twovec{p})$
in Eq.~(\ref{photoncorrelation}).
This approach is less likely to be sensitive to systematic errors
than the pole mass determination.

We note that the current operator $\Delta^{\xy}_{\vec{x},i}$, i.e.,
the dual of the photon,
has odd parity with respect to reflections of the gradient angle
$\theta_{\vec{x}+i}-\theta_{\vec{x}}$. The more standard compact functions
of the gradient angle 
\begin{eqnarray}
O_{o} (\vec{x},i)& = & i \beta^{-1} {\rm sin} (\theta_{\vec{x}+i}-\theta_{\vec{x}}) \\
O_{e} (\vec{x},i)& = &  ~\beta^{-1} {\rm cos} (\theta_{\vec{x}+i}-\theta_{\vec{x}})
\end{eqnarray}
have odd and even parity respectively. The correlation functions of these observables
map to correlation functions of 
\begin{eqnarray}
O_{o} (\vec{x},ij) & = & e^{-{\beta \over 2}}~ {\rm sinh}(\beta \plaq_{\vec{x},ij}  ) \\
O_{e} (\vec{x},ij) & = & e^{-{\beta \over 2}}~ {\rm cosh}(\beta \plaq_{\vec{x},ij}  )
\label{eq:dipole}
\end{eqnarray}
in the \fzs, 
again with definite parity.

\begin{figure}[tb]
\centerline{
 \psfig{file=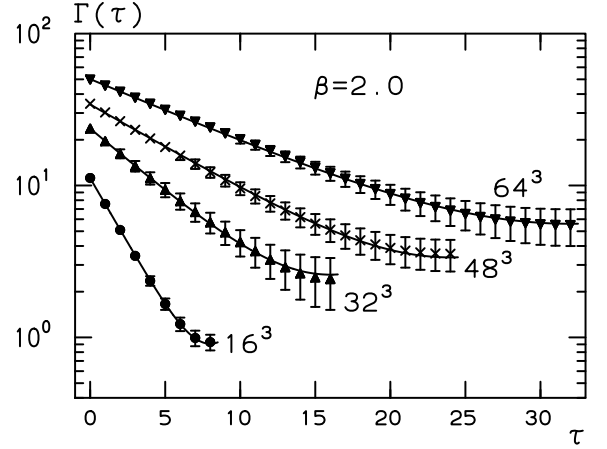,angle=270,width=8.5cm} 
}
\caption[a]{ 
The zero-momentum photon correlation 
function $\Gamma(\tau)$ at $\beta=2.0$ in the Coulomb
             phase. The fitted curves correspond to a massless
             photon. }
\label{fig:photon-mass-a}
\end{figure}

\begin{figure}
\centerline{
 \psfig{file=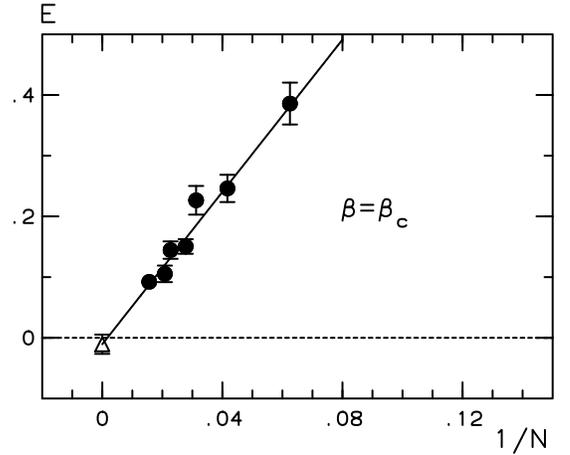,angle=270,width=8.5cm}
     }
\caption[a]{ Photon energy $E$, 
defined in Eq.~(\ref{eq:fit_form}), 
at $\beta=\beta_c$. 
             The straight line is a fit to Eq.~(\ref{equ:energyNfit}) 
             and extrapolates
             to vanishing photon mass at infinite volume.
           }
\label{fig:photon-mass-a2}
\end{figure}

We measure the photon mass from symmetric $N^3$ and elongated 
$N_3>N_2=N_1$ periodic lattices, using the lowest non-zero value 
of the momentum
\begin{equation}
\twovec{p}= \left(\frac{2 \pi}{N_1},0\right).
\label{eq:lowest_1_2_momentum}
\end{equation}
The asymptotic decay of the photon correlation function
$\Gamma(\tau,\twovec{p}) \sim e^{-E(\twovec{p})\tau}$
is governed by the lattice
dispersion relation
\begin{equation}
E(\twovec{p})= \sqrt{ 
                 {p^2_{\Lambda(1,2)}} + \hat{m}_\gamma^2 +\Sigma(\twovec{p})
                 },
\end{equation}
where
\begin{eqnarray}
 \hat{m}_\gamma &=& \frac{1}{2} {\rm sinh}(\frac{m_\gamma}{2}).
\label{equ:pLambda}
\end{eqnarray}
The photon self energy 
$\Sigma(\twovec{p})$ is expected to be small for large lattice sizes, and for
the data used in this work it is unobservable.

\begin{figure*}[tb]
\centerline{
 \psfig{file=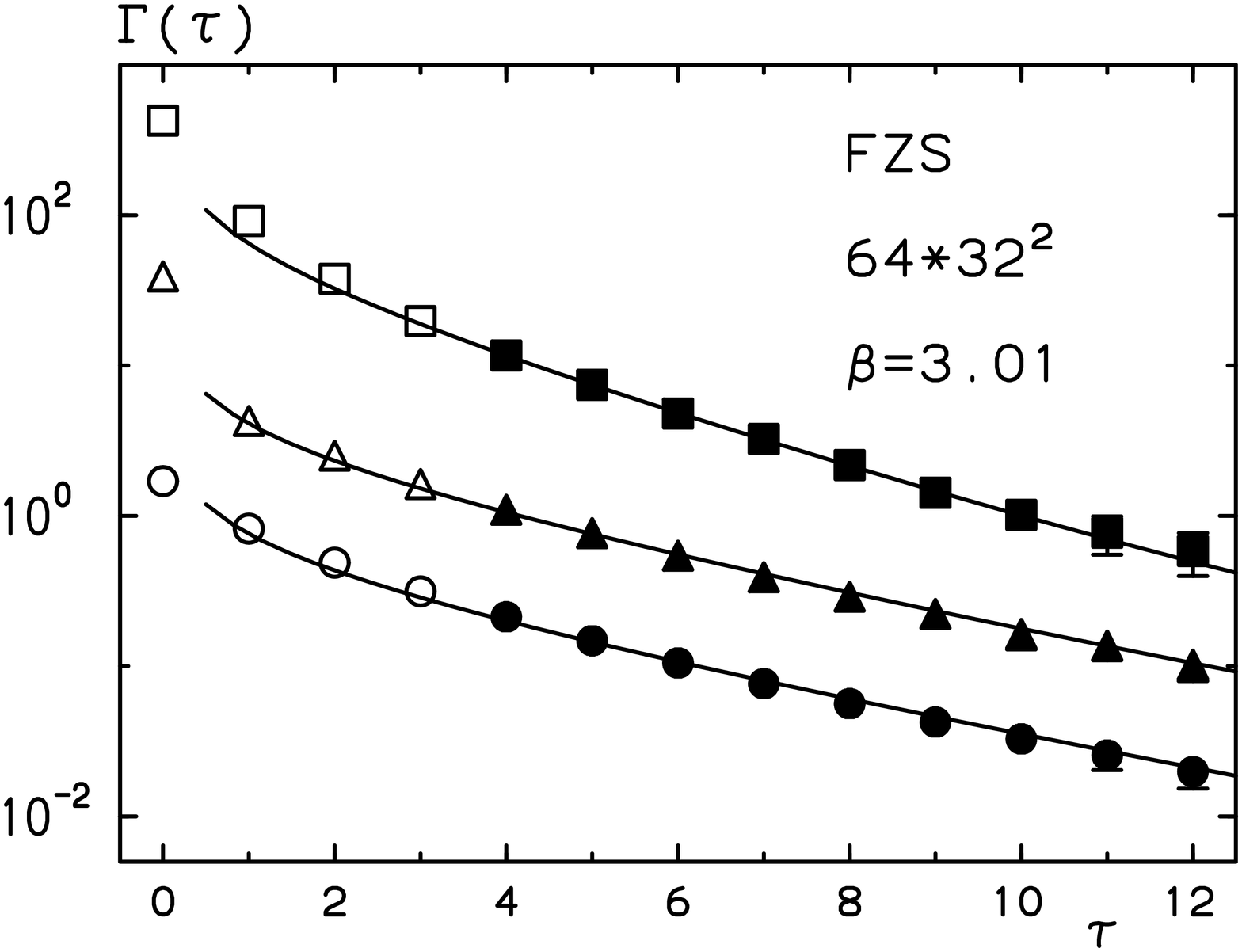,angle=270,width=7.5cm} 
 \psfig{file=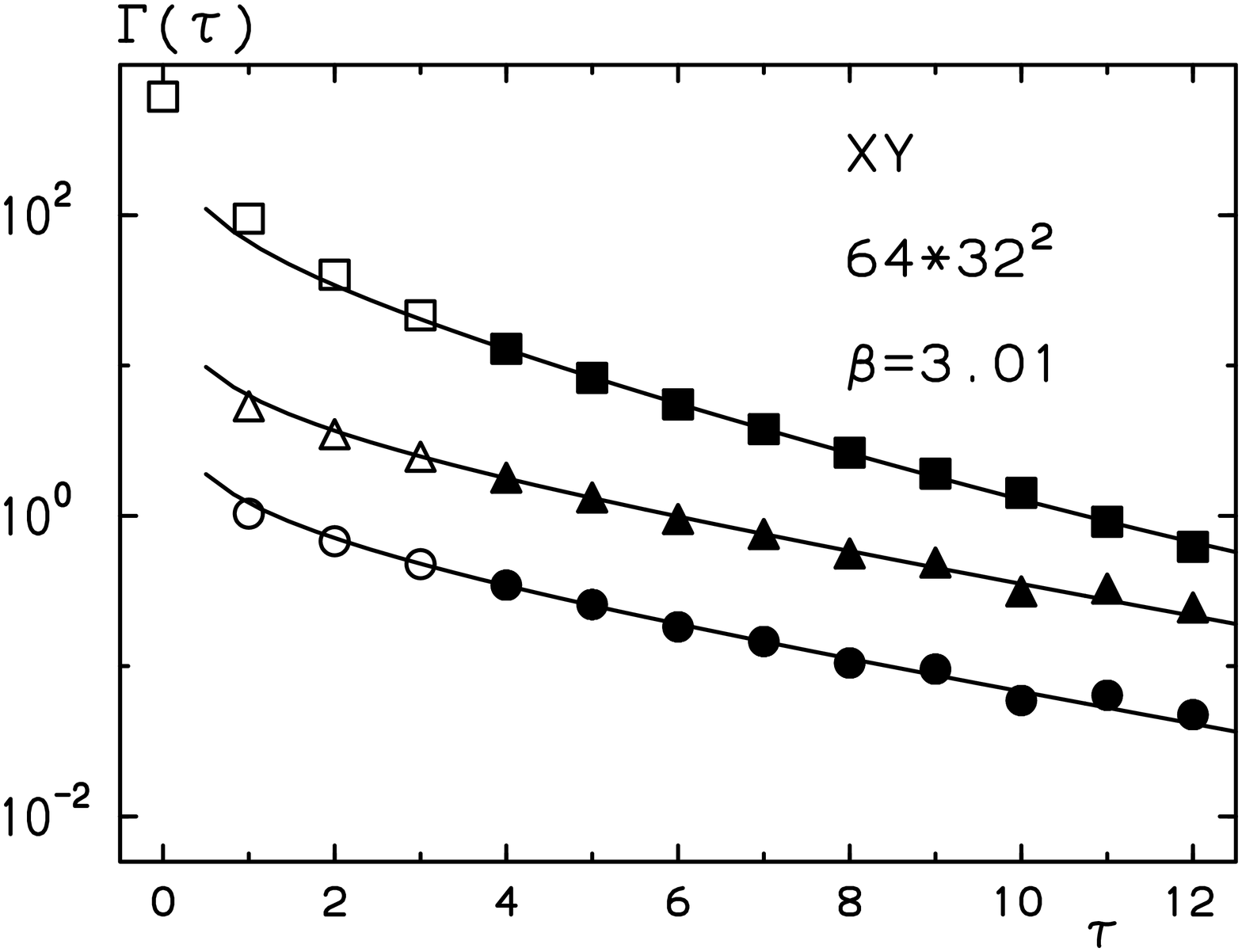,angle=270,width=7.5cm}
 }
\centerline{ \hfill (a) \hfill \hfill (b) \hfill}
\caption[a]{ (a) Correlation functions in the \fzs\ and
             (b) their dual
             counterpart
in the XY model.  The data is at $\beta=3.01$, in the 
             superconducting phase of the \fzs.  From bottom to top,
             the curves are the photon, $O_o$ and $O_e$ in the \fzs\ and
             their duals in the XY model.
           }
\label{fig:check-at-301}
\end{figure*}

The photon is massless in the Coulomb phase of the gauge theory. 
In Fig.~\ref{fig:photon-mass-a} we show the 
correlation functions $\Gamma(\tau)$ at $\beta=2$ measured 
from $16^3$ -- $64^3$ lattices.
The curves in the figure are fits to the data, assuming 
vanishing photon mass and no anomalous dimension,
\begin{eqnarray}
\Gamma(\tau)&=&A\left[ e^{-\tau\sqrt{2[1-{\rm cos}(2 \pi /N_1)]}}    
\right.\nonumber\\&&\left.
                ~+~e^{-(N_3-\tau)\sqrt{2[1-{\rm cos}(2 \pi /N_1)]}}   
              \right]~+~{\rm const},
\end{eqnarray}
which yields a perfect fit to the data.

The photon mass also
vanishes at the critical point $\beta=\beta_c$, but there one must
take into account the non-zero anomalous dimension
$\eta_A=1$. Instead of a pure exponential fit, 
we assume the asymptotic behaviour in Eq.~(\ref{equ:besselKasymp})
and use the ansatz
\begin{equation}
\Gamma(\tau)=A[ \tau^{-1/2}  e^{-\tau E}+
               (N_3-\tau)^{-1/2} e^{-(N_3-\tau)E}   
              ]~+~{\rm const}
\label{eq:fit_form}
\end{equation}
in our fits.

In Fig.~\ref{fig:photon-mass-a2}
we show the energy of the photon state with $|\twovec{p}| = 2\pi/N$, 
measured from cubical $N^3$ lattices, as a function of
$1/N$. Fitting the data with the form
\begin{equation}
E(N)=E(N=\infty)+A_E N^{-1},\quad \beta=\beta_c,
\label{equ:energyNfit}
\end{equation}
we obtain a vanishing photon energy $E(N=\infty)=0.01(2)$ 
and $A_E=6.3(6)$ with $\chi^2/\mbox{d.o.f.}$ value of $1.20$ for the fit.
The value of $A_E$ agrees very well with the value $2\pi$ expected if
$E(\twovec{p})=\sqrt{p_{\Lambda(1,2)}^2}$.

Let us now turn to the photon mass in the superconducting
phase.  At fixed $\beta=3.01$ we compare the finite-momentum 
photon correlation functions and correlation
functions of the observables $O_{e}$ and $O_o$ in the \fzs\ 
with their dual counterparts in the XY model. The simulations are
carried out
on $64 \times 32^2$ lattices with high
statistics of about $10^6$ sweeps.
Fig.~\ref{fig:check-at-301} displays three 
$|\twovec{p}| = 2\pi/32$
correlation functions for (a) the \fzs\ and (b) the XY model. 
The corresponding correlation functions
of the \fzs\ are not identical to the ones in the XY
model. Because the correlators have been measured between planes rather
than points, they are more sensitive to boundary conditions, and
therefore the discrepancy may well persist even in the
infinite-volume limit.

Again we use the ansatz in Eq.~(\ref{eq:fit_form}) to fit the data. 
The fits are done
in the $\tau$ interval $4 \le \tau \le 32$.  For the photon mass, 
we obtain the 
values $m_\gamma=0.08(3)$ in the {\fzs} and 
$m_\Delta=0.06(3)$ for its dual in the XY model. The values 
are consistent with each other and also with twice the XY scalar mass value 
$m(\beta=3.01)\approx 0.05(1)$, but the statistical errors are large.


Mass values from the odd parity $O_o$ 
correlation functions have the values $0.10(4)$ in the {\fzs}
and $0.06(3)$ in the XY model and therefore are degenerate
with the photon mass. For the even parity $O_e$ operator,
we obtain the mass $0.28(2)$ in the {\fzs} and $0.25(1)$
in the XY model.

\begin{figure}[tb]
\centerline{ 
   \psfig{file=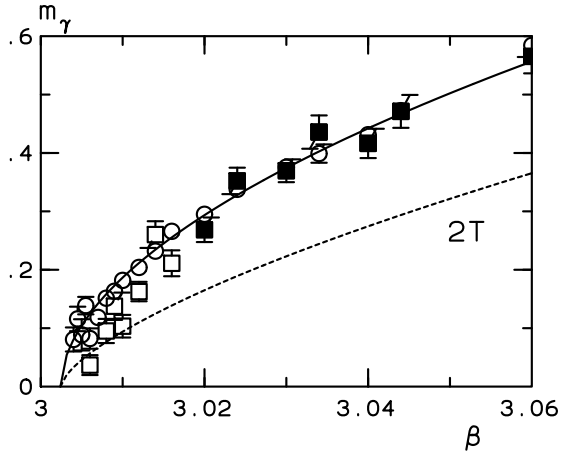,angle=270,width=9.2cm} 
}
\caption[a]{ The 
             photon mass $m_\gamma$ in the superconducting phase
             as a function of $\beta$. 
             The squares show the values determined 
	     from the exponential decay of the photon 
             correlation
             function, and the circles show the pole
             mass values from fits to Eq.~(\ref{ansatz}) in 
	     Sect.~\ref{sect:susceptibilities}.
             The solid line is a power law fit to 
             the
measured
             values of $m_\gamma$. (Only filled squares are used in
             the fit).             
             The dashed line shows the decay rate $m_\gamma=2\tens$,
             which corresponds to an unstable photon decaying into two
             vortices.
           }
\label{fig:photon-mass-b}
\end{figure}

In order to obtain the critical behaviour of the photon mass
$m_\gamma$ in the superconducting phase, we repeat the correlation
function analysis at several values of $\beta > \beta_c$, using
lattices of sizes $48^3$ and $64 \times 32^2$.
The final results are shown as circles in Fig.~\ref{fig:photon-mass-b}.
For comparison, the figure also shows $m_\gamma$
determined from the finite momentum dispersion 
of the gauge field susceptibility $\chi_A(\vec{p})$ in Section  
\ref{sect:susceptibilities}.  
Neglecting the points very close to $\beta_c$ 
and assuming a power law singular scaling behavior
\begin{equation}
m_\gamma(\beta) =  A_\gamma (\beta-\beta_c)^{\nu'} ,
\label{photon_scaling}
\end{equation}
a fit in the range $3.02 \le \beta \le 3.06$ to $m_\gamma$ from the 
correlation functions (squares) 
yields $A_\gamma=2.6(5)$ and $\nu'=0.54(6)$ with 
$\chi^2/\mbox{d.o.f.} = 1.3$ for the fit. 
This  value for the
exponent $\nu' \approx 1/2$ is inconsistent
with the XY value $\nu_{\rm XY} \approx 2/3$.  It agrees with 
the mean field value, as was predicted in Ref.~\onlinecite{kiometzis1994} 
and observed experimentally in Ref.~\onlinecite{paget1999}.

However, as we already argued in Section \ref{sect:susceptibilities}, the
scaling law \eq{photon_scaling} with $\nu' \approx 0.5 $ describes
only pre-asymptotic scaling, and the true critical exponent
$\nu'=\nu_{\rm XY}$.  As the measured values
of $m_\gamma$ are higher than $2\tens$, we expect the photon to decay
into two vortices, which would lead to $\nu'=\nu_{\rm XY}$, but also
to $m_\gamma=2\tens$.  This behaviour is shown in
Fig.~\ref{fig:photon-mass-b} by the dashed line, and is clearly
incompatible with the direct mass measurements except in very close
proximity of the critical point.  However, the statistical uncertainty
of the data in this region is too large to justify quantitative
comparison.  A possible source for the overall discrepancy is the fact
that both the lattice sizes and the distance where the mass is
extracted from the correlation functions is of order $1/m_\gamma$,
while in order to be able to neglect vortex interactions and observe
$m_\gamma \sim 2\tens$, distances and lattice
sizes much larger than this are required.  This is very
difficult to achieve in practice.

We also remark a further complication in the photon mass measurements.
Our simulations, which typically run for
about $10^6$ Monte Carlo sweeps, exhibit from time to time
``exceptional configurations'' with large contributions to the photon
correlator at large $\tau$-distances.  We suspect that excitations of
the vortex loop network are responsible.  Indeed, it is easy to see
that a vortex-antivortex pair wrapping around the finite lattice to
the $z$ direction also contributes to the photon correlation function
in Eq.~\nr{photoncorrelation}.  These configurations can significantly
affect the measurements, unless the lattice size is again much larger than
the inverse vortex tension.

\section{CONCLUSION}

In this paper, we have used numerical techniques to explore the
duality between the three-dimensional integer gauge theory and the XY
model in the Villain formulation. Our aim was to identify the ways in 
which the duality manifests itself in the correspondence between 
observables and the critical exponents of the two models.

First, we used the duality to test the method developed in
Ref.~\onlinecite{Kajantie:1999zn} for measuring the vortex tension. The
results agree perfectly with the direct measurements of the correlation
length in the XY model.  Similar methods can be used in (and were
originally developed for) more
complicated theories, which do not have an exact dual description.

Second, we investigated the critical behaviour of the magnetic field
of the gauge theory, which is dual to the Noether current of the XY
model. We measured the anomalous dimension $\eta_A$ at the critical
point, as well as the scaling exponents $\gamma_A$ and $\chimexp$ of
the gauge field susceptibility and the magnetic permeability. 
Our results are compatible with
predictions $\eta_A=1$,\cite{Bergerhoff:1995zq}
$\gamma_A=\nu$,\cite{olsson1998} and
$\chimexp=\nu$.\cite{fisher1973,olsson1998,son2002}

These results, together with the Fisher scaling
relation~(\ref{fishers_scaling}), imply that the photon mass (or the
penetration 
depth
 $\lambda$) scaling exponent
$\nu'$ must be equal to the XY model critical exponent:
$\nu'=\nu_{\rm XY}$.\cite{herbut1996} Our results
support this
prediction, but only in extremely close proximity to the critical point. 

The non-trivial anomalous dimension
makes the determination of the photon mass extremely difficult and
prone to systematic errors.
The ratio of the photon mass to the vortex
tension was ${A_{\gamma}/ A_{\tens}}\approx 3-4$ in the various
approaches we tried.  This means that the
photon ought to decay into two vortices and the photon correlator
should decay with rate $2\tens$ at long distances.
Our data seems compatible with this very close to the critical point,
but is not conclusive because of large statistical and systematical
uncertainties.

Nonetheless, we believe that the photon does indeed decay and the high
ratio ${A_{\gamma}/ A_{\tens}}$ is due to an incorrect ansatz for the
propagator.  Were the true form of the propagator in the
proximity of the critical point known, it ought to give ${A_{\gamma}/
A_{\tens}}=2$, at least if the finite volume effects can be avoided. 
This can be used as a 
stringent test for any theoretical
calculation of the propagator near the critical point.

The ratio ${A_{\gamma}/ A_{\tens}}$ can also be measured in
simulations of the full Ginzburg-Landau theory,\cite{inprogress} and
in principle also in superconductor experiments, since the
vortex tension is related to the critical field strength by
$H_{c1}={\cal T}/2 \pi$.
We believe that such experiments would face the same difficulties in
measuring the penetration 
depth
 as we did with the photon mass, but
if a reliable measurement can be carried out, the ratio
$\lambda^{-1}/H_{c1}$ could be used to estimate
how much closer to the critical point one would
have to go to see the true scaling in the penetration
depth.
As in the experiments in Ref.~\onlinecite{paget1999}, the 
penetration
depth
 $\lambda=m_\gamma^{-1}$ we measured in our simulations
was apparently obeying the mean-field scaling $\nu'=1/2$, but it
became consistent with the inverted XY behaviour ${A_{\gamma}/
A_{\tens}}=2$ when $(\beta-\beta_c)/\beta_c\lsim 0.002$.

\section*{Acknowledgements}
We would like to thank K.~Kajantie, H.~Kleinert, M.~Laine and  D.~Litim
for useful
discussions. AR was supported by PPARC and also in part by 
the ESF COSLAB programme and the National Science Foundation 
Grant No. PHY99-07949.

\begin{widetext}
\appendix
\section{Current correlator in free-field theory}
\label{sect:app}

In this appendix we derive Eq.~(\ref{equ:arctan}) by assuming that 
the vortices are non-interacting free fields at the critical
point. This means that the dual theory is simply a continuum
theory of a free
complex scalar field $\phi$ with mass $m=\tens$.

The current operator $j_i(\vec{x})$ is defined as
$j_i={\rm Im} \phi^*\partial_i\phi$. We want to calculate
the current-current correlator 
$
\langle j_i(-\vec{p})
j_j(\vec{p})\rangle$.
To do this, we write
\begin{equation}
j_i(\vec{p})
=
\int d^3x e^{i\vec{p}\cdot\vec{x}}j_i(x)=
-{1\over 2}\int {d^3q\over (2\pi)^3}
(2q_i+p_i)
\phi^*(\vec{p}+\vec{q})\phi(\vec{q}).
\end{equation}
Then the correlator is
\begin{equation}
\langle j_i(\vec{p})
j_j(\vec{p}')\rangle
=
{1\over 4}
\int {d^3q\over (2\pi)^3}{d^3r\over (2\pi)^3}
(2q_i+p_i)(2r_j+p'_j)
\langle
\phi^*(\vec{p}+\vec{q})\phi(\vec{q})\phi^*(\vec{p}'+\vec{r})\phi(\vec{r})
\rangle.
\end{equation}
Assuming that the vortices do not interact, the expectation value
factorizes,
\begin{equation}
\langle j_i(\vec{p})
j_j(\vec{p}')\rangle
=
{1\over 4}
\int {d^3q\over (2\pi)^3}{d^3r\over (2\pi)^3}
(2q_i+p_i)(2r_j+p'_j)
\langle
\phi^*(\vec{p}+\vec{q})
\phi(\vec{r})
\rangle
\langle
\phi^*(\vec{p}'+\vec{r})
\phi(\vec{q})
\rangle,
\end{equation}
and using the tree-level propagator
\begin{equation}
\langle
\phi^*(\vec{p})
\phi(\vec{p}')
\rangle
=(2\pi)^3\delta(\vec{p}-\vec{p}'){1\over p^2+m^2},
\end{equation}
we obtain the one-loop integral
\begin{equation}
\langle j_i(\vec{p})
j_j(\vec{p}')\rangle
=(2\pi)^3\delta(\vec{p}+\vec{p}')
{1\over 4}
\int {d^3q\over (2\pi)^3}
{(2q_i+p_i)(2q_j+p_j) \over (q^2+m^2)
((\vec{p}+\vec{q})^2+m^2)}.
\end{equation}
Now, this is a tensor that depends only on one vector $\vec{p}$, so it
must be generally of the form
\begin{equation}
\langle j_i(-\vec{p})
j_j(\vec{p})\rangle
=
A(p)\Big(\delta_{ij}-{p_ip_j\over p^2}\Big)
+B(p){p_ip_j\over p^2},
\end{equation}
where $A(p)$ and $B(p)$ are functions of the absolute value of
$\vec{p}$ only.
The duality in 
Eq.~(\ref{equ:dual2point})
implies that up to a constant multiplicative factor and with some
constant $C$,
the photon correlator is 
\begin{equation}
\langle B_i(-\vec{p})
B_j(\vec{p})\rangle
=\Big[A(p)+C\Big]\Big(\delta_{ij}-{p_ip_j\over p^2}\Big)
+\Big[B(p)+C\Big]{p_ip_j\over p^2}
.
\end{equation}
Since we know that $B_i$ is sourceless, we must have $B(p)=-C$, and
it is straightforward to calculate its value
\begin{equation}
B(p)={p_ip_j\over p^2}\langle j_i(-\vec{p})
j_j(\vec{p})\rangle
=
{1\over 4p^2}\int {d^3q\over (2\pi)^3}
{(2\vec{p}\cdot\vec{q}+p^2)^2 \over(q^2+m^2)
((\vec{p}+\vec{q})^2+m^2)}=
-{m\over 8\pi}.
\end{equation}
This means that
the gauge field susceptibility $\chi_A(p)$
defined in Eq.~(\ref{equ:chiAp}) is
\begin{equation}
\chi_A(p)={1\over p_x^2+p_y^2}\langle B_3(-\vec{p})
B_3(\vec{p})\rangle
={1\over p_x^2+p_y^2}
\left(1-{p_z^2\over p^2}\right)\left[A(p)+\frac{m}{8\pi}\right]
={1\over p^2}\left[A(p)+\frac{m}{8\pi}\right].
\end{equation}
\end{widetext}

We can calculate $A(p)$ by contracting $
\langle j_i(-\vec{p})
j_j(\vec{p})\rangle$ with the transverse projection operator,
\begin{equation}
(D-1)A(p)={1\over p^2}\int {d^3q\over (2\pi)^3}
{p^2q^2-(\vec{p}\cdot\vec{q})^2 \over (q^2+m^2)
((\vec{p}+\vec{q})^2+m^2)},
\end{equation}
and calculating the integral, we find
\begin{equation}
A(p)=-{m\over 16\pi}+{p^2-4m^2\over 32\pi p}\arctan{p\over 2m},
\end{equation}
and consequently
\begin{equation}
\chi_A(p)={1\over 16\pi p^2}
\left[ m + {p^2-4m^2\over 2p}\arctan {p\over 2m}\right],
\end{equation}
up to an overall constant factor, which we parameterize by $c_A$ in
Eq.~(\ref{equ:arctan}). The value of $m_\gamma$, on the other hand, is
fixed to $m_\gamma=2m=2\tens$.

\end{document}